\documentclass[aps,prb,twocolumn,superscriptaddress,showpacs,amssymb,amsmath,floatfix]{revtex4-1}

\usepackage{graphicx,color}
\usepackage{color,verbatim}
\usepackage{hyperref}

\newcommand{\A}{\ensuremath{\mathbf{A}}}

\newcommand{\bl}{\ensuremath{\lambda}}

\newcommand{\e}{\ensuremath{\varepsilon}}
\newcommand{\g}{\ensuremath{\mathbf{g}}}
\newcommand{\G}{\ensuremath{\mathbf{G}}}
\newcommand{\Ga}{\ensuremath{\mathbf{\Gamma}}}
\newcommand{\Ha}{\ensuremath{H}}
\newcommand{\Ht}{\ensuremath{\mathcal{H}_{\e'}}}

\newcommand{\K}{\mathbf{K}}
\newcommand{\llangle}{\langle\!\langle}
\newcommand{\M}{\ensuremath{\mathbf{M}}}
\newcommand{\np}{\ensuremath{\mathcal{N}_0}}
\newcommand{\n}{\ensuremath{\mathbf{n}}}
\newcommand{\om}{\ensuremath{\omega}}

\newcommand{\Q}{\ensuremath{\mathbf{Q}}}

\newcommand{\rrangle}{\rangle\!\rangle}
\newcommand{\Si}{\ensuremath{\mathbf{\Sigma}}}
\newcommand{\T}{\ensuremath{\mathcal{T}}}
\newcommand{\Te}{\tilde{\mathbf{T}}}
\newcommand{\Tlr}{\mathbf{T}}
\newcommand{\Tr}[1]{{\rm Tr}\{ #1\}}

\begin{document}

\title{Current noise in molecular junctions: effects of the electron-phonon interaction}

\author{Federica Haupt}\email{federica.haupt@uni-konstanz.de}
\affiliation{Fachbereich Physik, Universit\"at Konstanz, D- 78457
Konstanz, Germany}

\author{Tom\'{a}\v{s} Novotn\'y}\email{tno@karlov.mff.cuni.cz}
\affiliation{Department of Condensed Matter Physics, Faculty of
Mathematics and Physics, Charles University in Prague, Ke Karlovu 5,
CZ-121 16 Praha 2, Czech Republic}

\author{Wolfgang Belzig}\email{wolfgang.belzig@uni-konstanz.de}
\affiliation{Fachbereich Physik, Universit\"at Konstanz, D- 78457
Konstanz, Germany}

\begin{abstract}
We study inelastic effects on the electronic current noise in molecular junctions, due to the coupling between transport electrons and vibrational degrees of freedom. Using a full counting statistics approach based on the generalized Keldysh Green's function technique, we calculate in an unified manner both the mean current and the zero-frequency current noise. For multilevel junctions with weak electron-phonon coupling, we give analytical formulas for the lowest order inelastic corrections to the noise in terms of universal temperature- and voltage-dependent functions and junction-dependent prefactors, which  can be evaluated microscopically, e.g. with {\em ab-initio} methodologies. We identify distinct terms corresponding to the mean-field contribution to noise and to the  vertex corrections, and we show that the latter contribute substantially to the inelastic noise. Finally, we illustrate our results by a simple model of two electronic levels which are mutually coupled by the electron-phonon interaction and show that the inelastic noise spectroscopy is a sensitive diagnostic tool.                               
\end{abstract}

\date{\today}
 \maketitle

\section{Introduction}
Recent progress in experimental techniques, such as break junctions and scanning tunneling microscopy, allows to electrically contact single molecules,  to create and manipulate atomic wires, and to investigate the electronic transport properties of these nanoscopic objects.\cite{Cuniberti, Agrait:PhysRep} Contrary to larger devices, atomic-scale ones usually retain their microscopic features, which are then observable in the transport spectroscopy. Apart from the purely electronic contributions, other degrees of freedom such as vibrational modes or local spins can be addressed and revealed by  point-contact spectroscopy\cite{Naidyuk} (PCS) or by inelastic electron tunneling spectroscopy (IETS).\cite{Jaklevic}
These techniques have been extensively used in the recent past to reveal inelastic features in the non-linear conductance due to vibrations\cite{Stipe, Agrait:PRL02, Smit, Djukic_PRB, Tal, Franke,Rahimi:2009, Arroyo:2010} or local spin excitations,\cite{Eigler,Heinrich,Cinane} triggering an intense theoretical activity.  So far most studies focused on the current-voltage characteristics, and present day PCS/IETS theories,\cite{LorentePersson, Frederiksen:PRL04, Paulsson:RapCom, Viljas, delaVega,Solomon,  Sergueev:2007,Frederiksen:PRB07,Teobaldi:2007, Frederiksen:2007, Paulsson:PRL100, Iben, Alducin:PRL10,Fransson:NL10,Monturet:2010,Patton} often based on ab-initio calculations, allow to make detailed predictions for the conductance that compare favorably with experimental results. 

Electronic current (shot) noise\cite{Beenakker,Blanter} is another quantity of fundamental interest in transport, representing the second cumulant of the current distribution within the full counting statistics methodology.\cite{Nazarov:AnnPhys, *Nazarov:book} Although the measurement of higher order cumulants is experimentally challenging, shot noise in atomic contacts and molecular junctions has been already measured in the small voltage (elastic) regime,\cite{VandenBrom,Djukic,Kiguchi} and there are ongoing experimental efforts to address the inelastic noise signal as well.\footnote{J.~M.~van Ruitenbeek, private communication.}  These progresses made the investigation of effects due to electron-phonon ($e$-ph) interaction on the current noise an appealing task from the theoretical point of view, with the ultimate goal of allowing for prediction for the noise in molecular junctions as reliable as those now available for the non-linear conductance.

Since the noise is technically represented by a two-particle non-equilibrium correlation function, its direct evaluation poses a significant challenge compared to the mean current.  For molecular junctions weakly coupled to leads, noise calculations based on one-level models have been put forward within the rate equation approach.\cite{Mitra,KochPRL94,*KochPRL95} In the opposite limit, pioneering works based on the non-equilibrium Green's functions formalism\cite{Zhu,Galperin:PRB06} adopted a mean-field-like approximation for the noise, thus neglecting the contributions due to the  vertex corrections. A very convenient tool to overcome these limitations is the full counting statistics,\cite{Nazarov:AnnPhys, *Nazarov:book, Lesovik, *LLL:JMP} since it allows to address the noise and other current cumulants, while taking consistently into account all contributions due to $e$-ph coupling up to a given order in perturbation theory.
Simultaneously with other two groups,\cite{Schmidt,Avriller} we exploited such an approach
to analyze the transport properties of a simple model for molecular junctions consisting of a single resonant level symmetrically 
coupled to metallic leads and weakly interacting with a single phonon mode.\cite{Haupt} Despite its simplicity, this model can be applied to the experimentally relevant case of junctions formed by a single hydrogen/deuterium molecule between platinum leads,\cite{Smit} and in this case we predicted a significant inelastic contribution to the current noise.\cite{Haupt} 

In this paper we go beyond such a simple model and we extend our theory for inelastic noise\cite{Haupt} to more complex molecular junctions and to atomic wires. In fact,  we consider  the case of a junction formed by multiple electronic levels with asymmetric coupling to leads, and derive analytical formulas for the corrections to current and noise due to weak electron-phonon coupling. We express our result in terms of universal temperature- and voltage-dependent functions and junction-dependent prefactors.  These expressions, when supplemented with {\em ab-initio} calculations to estimate microscopically the prefactors characterizing a given junction, can serve as a basis to make realistic predictions for the current noise in a relevant class of molecular and atomic-size junctions. In this respect, our work  can be viewed as a direct extension of the corresponding lowest order expansion scheme developed for the non-linear conductance.\cite{Paulsson:RapCom,Viljas,delaVega}  In addition, we identify the contributions to noise due to the vertex corrections and show that, even in limit of weak $e$-ph coupling, they need to be taken into account in order to obtain accurate results and to comply with the fluctuation-dissipation theorem.

The paper is organized as follows. After a brief description of the model of a multi-level junction coupled to leads and weakly interacting with a number of vibronic modes (phonons) in Sec.~\ref{Sec:model}, we introduce the methodology of the noise calculation via extended Keldysh Green's functions in  Sec.~\ref{Sec:methods}. In Sec.~\ref{Sec:elastic} we consider the case of no interactions and recover the standard results for the elastic current and noise. Our original contribution is presented in Secs.~\ref{Sec:Hartree} and  \ref{Sec:Fock}, where we discuss the corrections to the current and noise due to the Hartree and the Fock diagrams, respectively.  
In subsection \ref{Subsec:indeplev} we then illustrate our theory by a simple example of independent electronic levels which are coupled only via the $e$-ph interaction. Finally, we conclude and give an outlook of open issues and possible extensions of the present work in Sec.~\ref{Sec:concl}. More technical parts of the text are deferred to 5 appendices.  
In addition, we make use of the Electronic Physics Auxiliary Publication Service (EPAPS) to supplement the paper with a \texttt{Mathematica} notebook with full expressions for the lowest order corrections to current and noise due to $e$-ph coupling. This file is intended to be of use for interested readers in order to access directly our results without need of retyping cumbersome formulas from the main text, and it also extends the results of subsection \ref{sec:SF} to the case of finite temperature.                                      


\section{Model}
\label{Sec:model}
The system we consider can be
schematically represented as a central device region (representing the molecule or the atomic-wire) which is
tunnel-coupled to non-interacting metallic leads
\begin{equation}\label{eq:H}
\hat{\Ha}=\hat{\Ha}_C+\hat{\Ha}_{L,R}+\hat{\Ha}_T.
\end{equation}
Neglecting for simplicity the spin degree of freedom\footnote{Based on a spin-less model, our results need to be multiplied by a factor of 2 when compared with works where spin degeneracy is explicitly taken into account.}
the central region can be described by the following Hamiltonian
\begin{subequations}
\begin{align}
\hat{\Ha}_C &=\hat{\Ha}_{\rm d}+\hat{\Ha}_{\rm ph}+\hat{\Ha }_{e\rm ph}\\
\hat{\Ha}_{\rm d}&=\sum_{i,j}h_{\rm d}^{ij}\hat{d}^{\dag}_{i}\hat{d}_j\\
\hat{\Ha}_{\rm ph} &=\sum_{\ell}\hbar\om_{\ell}\hat{b}_{\ell}^{\dag}\hat{b}_{\ell}\\
\hat{\Ha }_{e\rm ph} &=\sum_{\ell}\sum_{i,j}M_{\ell}^{ij}\hat{d}^{\dag}_i\hat{d}_j(\hat{b}_{\ell}^{\dag}+\hat{b}_{\ell}),
\end{align}
\end{subequations}
where $\hat{d}_i$ ($\hat{d}^{\dag}_i$) and $\hat{b}_{\ell}$
($\hat{b}^{\dag}_{\ell}$) are the electron and phonon annihilation
(creation) operators, respectively; $\hat{\Ha}_{\rm d}$ is the single-particle effective
Hamiltonian of the electrons moving in a
static arrangement of atomic nuclei, $\hat{\Ha}_{\rm ph} $ is the Hamiltonian
of free uncoupled phonons, $\hat{\Ha }_{e\rm ph}$ is the $e$-ph
coupling within the harmonic approximation, and $\M_{\ell}=\{ M_{\ell}^{ij}\}$ is the $e$-ph
coupling matrix for the ${\ell}$-th phonon mode. Here, boldface
notation stands for matrices in the system electronic space.  The leads and
tunneling Hamiltonians are given by
\begin{align}
\hat{\Ha}_{L,R}&=\sum_{k,\alpha=L,R}\varepsilon_{\alpha,k}\hat{c}^{\dag}_{\alpha,k}\hat{c}_{\alpha,k},\\
\hat{\Ha}_T&=\sum_{k,\alpha=L,R}\sum_{i}(V_{\alpha,k}^i\hat{c}^{\dag}_{\alpha,k}\hat{d}_i^{\phantom{\dagger}}+
h.c.).
\end{align}
The states in the leads are occupied according to the Fermi
distributions $f_{\alpha}(\e)=f(\e-\mu_{\alpha})$, with
$f(\e)=(1+e^{\beta \e})^{-1}$,  $\beta=1/k_B T$ the inverse
temperature, and $\mu_ {\alpha}$ the chemical potential of
lead-$\alpha$. The applied bias voltage is $eV=\mu_L-\mu_R$.

\section{Methods}
\label{Sec:methods}
\subsection{The generalized Keldysh Green's function technique}
\label{Subsec:Keldysh}
To calculate the average current and the zero-frequency noise in the stationary regime, we employ the generalized
non-equilibrium Keldysh Green's function technique.~\cite{Nazarov:AnnPhys,*Nazarov:book}
In this approach, one is interested in finding the cumulant
generating function $\mathcal{S}(\lambda)$, which in the case of
charge transport is defined as
\begin{equation}
e^{-\mathcal{S}(\lambda)}=\sum_{N}e^{i N\lambda}P_{t_{0}}(N)
\end{equation}
where $P_{t_{0}}(N)$ is the probability for $N$ charges to be transferred through the
system during the measuring time $t_{0}$ and $\lambda$ is a continuous parameter known
as {\em counting field}.  Given $\mathcal{S}(\lambda)$, the cumulants $\llangle  \delta N^{k}  \rrangle $ of the charge
transfer distribution can be straightforwardly calculated according to the prescription
\begin{equation}
\llangle \delta N^{k} \rrangle =-\left. \frac{\partial^{k}}{\partial (i\lambda)^{k}}\mathcal{S}(\lambda)\right|_{\lambda=0}.
\end{equation}
Under the assumption that the measuring time $t_{0}$ is much longer than any correlation time of the
system ($t_{0}\to \infty$), the first two cumulants of $P_{t_{0}}(N)$  are directly proportional to the
average current through the system $I$ and to the zero-frequency current noise $S$,
\begin{equation} \label{eq:IS}
I=e\frac{\llangle \delta N\rrangle}{t_{0}},\qquad S=e^{2}\frac{\llangle \delta N^{2}\rrangle}{t_{0}},
\end{equation}
which are the quantities we are primarily interested in.

The key idea for calculating the cumulant generating function for
transport through a quantum system is to modify the Hamiltonian by
introducing a time-dependent counting field $\lambda(t)$ and to
relate $\mathcal{S}(\lambda)$ to the Keldysh Green's function of
the system in the presence of $\bl(t)$, i.e.\ to
$G_{\bl}^{ij}(t,t^{\prime})=-i\hbar^{-1}\langle {\cal
  T_{C}} \hat{d}_{i}(t) \hat{d}_{j}^{\dagger}(t^{\prime})\rangle_{\lambda}$, where the
evolution is due to the modified Hamiltonian.~\cite{Lesovik, *LLL:JMP, Nazarov:AnnPhys,*Nazarov:book} One way to accomplish this is to add a
time-dependent phase $\lambda(t)/2$ to the tunneling
matrix elements $V^{i}_{L,k}$,
\begin{equation*}
\hat{\Ha}_T\to\hat{\Ha}_T^{\lambda}=\sum_{k,j}V_{L,k}^j e^{-i\lambda(t)/2}\hat{c}^{\dag}_{L,k}\hat{d}_j^{\phantom{\dagger}}+V_{R,k}^j \hat{c}^{\dag}_{R,k}\hat{d}_j^{\phantom{\dagger}}+
h.c.
\end{equation*}
with $\bl(t)=\lambda\, \theta(t)\theta(t_{0}-t) $ on the forward branch of the Keldysh contour and $\bl(t)=-\lambda\, \theta(t)\theta(t_{0}-t) $ on the backward one,~\cite{Levitov}
where $\theta(x)$ is the Heaviside step-function.

Here we extend the result derived by Gogolin and
Komnik~\cite{GogolinKomnik} for the Anderson model to the case in
which the central region has several electronic states.
Generalizing the derivation of Ref.~\onlinecite {GogolinKomnik} to
a multilevel system, we obtain the following expression for the derivative of the
cumulant generating function
\begin{equation}\label{eq:CGF}
\frac{\partial{\mathcal{S}}( \bl)}{\partial{\lambda}}=t_{0} \int
\frac{d \e}{2 \pi \hbar} {\rm Tr}_{K}\left\{ \check{\Si}_{T}'(\e)\check{\G}_{\bl}(\e) \right\}
\end{equation}
where $\check{\G}_{\bl} $ represents the Keldysh-Green's function of
the system in Keldysh space
\begin{equation}
\check{\G}_{\bl}(\e) =\left(
\begin{array}{cc}
\G^{--}_{\bl}(\e)&\G^{-+}_{\bl}(\e)\\
\G^{+-}_{\bl}(\e)&\G^{++}_{\bl}(\e)
\end{array}
\right),
\end{equation}
$\check{\Si}_{T}'\equiv\partial \check{\Si}_{T}/\partial{\lambda}$, with $\check{\Si}_{T}$  the self-energy due to the modified tunneling Hamiltonian $\hat{\Ha}_{T}^{\bl}$
\begin{widetext}
\begin{equation}\label{eq:sigmaT}
  \check{\Si}_{T}(\e)=\left(
  \begin{array}{cc} i\sum_{\alpha=L,R}\Ga_{\alpha}[f_{\alpha}(\e)-1/2]&   -i\Ga_Le^{i\bl}f_L(\e)-i \Ga_R f_R(\e)\\
                             i \Ga_Le^{-i\bl}[1-f_L(\e)]+i\Ga_R[1-f_R(\e)]& i\sum_{\alpha=L,R}\Ga_{\alpha}[f_{\alpha}(\e)-1/2]
  \end{array}
  \right),
\end{equation}
\end{widetext}
and ${\rm Tr}_{K}$ stands for the trace over the electronic
degrees of freedom {\em and}  the Keldysh space, i.e. ${\rm
Tr}_{K}\{ \check{\mathbf O} \}={\rm Tr}\{ {\mathbf
O}^{--}+{\mathbf O}^{++} \}$, with ${\rm Tr}\{\cdots\}$ being the trace
in the system electronic space.  The check sign $\check{\phantom{o}}$ indicates matrices in the Keldysh space and the superscripts $-/+$ correspond to the
forward/backward branch of the Keldysh-contour. Note that in
Eq.~\eqref{eq:sigmaT} we have used the following sign convention for
the elements of the Keldysh-matrix for the self-energy
$\check{\Si}$
\begin{equation}
\check{\Si} =\left(
\begin{array}{cc}
\Si^{--}&-\Si^{-+}\\
-\Si^{+-}&\Si^{++}
\end{array}
\right).
\end{equation}
Finally,  $\Gamma_{\alpha}^{ij}(\e)=2 \pi \sum_k V^i_{\alpha,k}
{V^{{j}^*}_{\alpha,k}}\delta(\e-\e_{k,\alpha})$ is the level
broadening due to the coupling to the lead $\alpha$.

According to Eq.~\eqref{eq:CGF}, the problem of evaluating current
and noise (as well as any higher order cumulant of the charge
transfer distribution) is reduced to the calculation of the system single-particle
Green's function $\check{\G}_{\bl}$. The latter can be obtained
from the solution of the Dyson equation
\begin{equation}
\check{\G}_{\lambda}(\e)={\check{\g}}_{\lambda}(\e)+{\check{\g}}_{\lambda}(\e)\check{\Si}_{e\rm ph}(\e){\check{\G}}_{\lambda}(\e),
\end{equation}
where $\check{\Si}_{e\rm ph}$ is  the self-energy solely due to the
{\em e}-ph coupling, and $\check{\g}_{\bl}$ is the free Green's
function of the system in the presence of the leads and of the
counting field but without the {\em e}-ph interaction $\check{\g}_{\bl}=(\check{\g}_{\rm d}^{-1}-\check{\Si}_{T})^{-1}$, with
\begin{equation}
\check{\g}_{\rm d}(\e) =\left(
\begin{array}{cc}
\e\mathbf{1}-\mathbf{h}_{\rm d}&0\\
0&-\e\mathbf{1}+\mathbf{h}_{\rm d}
\end{array}
\right)^{-1},
\end{equation}
the Green's function of the isolated dot. 
It is important to notice that $\check{\Si}_{e\rm ph}$,
 depending on the Green's function of the system,
is itself a function of the counting field $\bl$ (see
Sec.~\ref{sec:HartreeFock}).

Finally, we remark in passing that for $\bl \neq 0$  it is
$\g^{--}_{\bl}+\g^{++}_{\bl}\neq \g^{-+}_{\bl}+\g^{+-}_{\bl}$,
i.e. in the presence of the counting field $\bl$, the four
Keldysh Green's functions are all independent.

\subsection{Current and Noise}
\label{Subsec:curnoi}
Although Eq.~\eqref{eq:CGF} gives access to  all  cumulants of the charge transfer distribution through the system,
in this work we will focus only on the study of the average current $I$ and the zero frequency noise $S$,
which are the quantities most easily accessible from the experimental point of view.

The average current is directly obtained from Eq.~\eqref{eq:CGF} by setting $\bl=0$
\begin{equation}\label{eq:I}
I=i e \int
\frac{d \e}{2 \pi \hbar} {\rm Tr}_{K}\left\{ \check{\Si}_{T}'(\e)\check{\G}_{\bl}(\e) \right\}_{\bl=0},
\end{equation}
while the noise is given by
\begin{equation}\label{eq:S}
\begin{split}
S =&\phantom{+} e^{2} \int \frac{d \e}{2 \pi \hbar} {\rm
Tr}_{K}\left\{ \check{\Si}_{T}''\check{\G}_{\bl}
+\check{\Si}_{T}' \check{\G}_{\bl}\check{\Si}_{T}' \check{\G}_{\bl}   \right\}_{\bl=0}\\
&+e^{2} \int
\frac{d \e}{2 \pi \hbar} {\rm Tr}_{K}\left\{ \check{\Si}_{T}' \check{\G}_{\bl}\check{\Si}_{e\rm ph}' \check{\G}_{\bl}   \right\}_{\bl=0},
\end{split}
\end{equation}
where, we have used the identity $\partial_{\bl} \check{\mathbf
O}=-\check{\mathbf O} (\partial_{\bl} \check{\mathbf
O}^{-1})\check{\mathbf O}$ together with the Dyson equation
$\check{\mathbf G}_{\bl}^{-1}=\check{\g}_{\rm
d}^{-1}-\check{\Si}_{T}(\bl)-\check{\Si}_{e\rm ph}(\bl)$. It turns
out that the first term of Eq.~\eqref{eq:S} corresponds exactly to
Eq.~(30) of Ref.~\onlinecite{Souza}, which gives the expression
for the noise within a mean-field approximation (see
Appendix ~\ref{App:MeanField}). For this reason, we identify
\begin{equation}\label {eq:Smf}
S^{\rm (mf)}=e^{2} \int
\frac{d \e}{2 \pi \hbar} {\rm Tr}_{K}\left\{ \check{\Si}_{T}''\check{\G}_{\bl}
+\check{\Si}_{T}' \check{\G}_{\bl}\check{\Si}_{T}' \check{\G}_{\bl}   \right\}_{\bl=0}
\end{equation}
as the {\em mean-field contribution} to noise. The remaining term constitutes the {\em vertex correction}
\begin{equation}\label{eq:Svc}
S^{\rm (vc)}=e^{2} \int
\frac{d \e}{2 \pi \hbar} {\rm Tr}_{K}\left\{ \check{\Si}_{T}' \check{\G}_{\bl}\check{\Si}_{e\rm ph}' \check{\G}_{\bl}   \right\}_{\bl=0}.
\end{equation}
As we will discuss in detail in the following, the vertex
correction $S^{\rm (vc)}$ can give a significant contribution to the total
noise, comparable to the mean-field part and thus, contrary to what was done in some pioneering works,\cite{Zhu, Galperin:PRB06}  it cannot be omitted even in the limit of weak interaction. Moreover, neglecting $S^{\rm (vc)}$ generally leads to violation of the fluctuation-dissipation theorem, see Appendix ~\ref{App:FDT}.

\subsection{Weak electron-phonon coupling}\label{sec:HartreeFock}
\label{Subsec:LOE}

\begin{figure}
\begin{center}
{\resizebox{0.8\columnwidth}{!}{\includegraphics{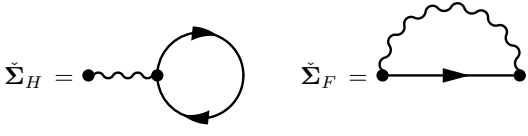}}}
\end{center}
\caption{Diagrammatic representations of $\check{\Si}_{H}$  and
$\check{\Si}_{F}$. Here,  the plain line stands for the free electronic Green's function $\check{\g}_{\bl}$, and the wiggly line for the phononic one $\check{d}_{\ell}$. Finally, the dot represents the $e$-ph coupling constant $\M_{\ell}$.  }\label{fig:diagrams:HF}
\end{figure}

In order to make use of Eqs.~\eqref{eq:I}, \eqref{eq:S}, we need
to determine the full Green's function $\check{\G}_{\lambda}$.
Being interested in the experimentally relevant limit of {\em
weak} electron-phonon coupling, we truncate the Dyson equation at
the lowest (second) order  in the {\em e}-ph coupling
\begin{equation}\label {eq:Gapprox}
\check{\G}_{\lambda}\approx{\check{\g}}_{\lambda}+{\check{\g}}_{\lambda}\check{\Si}^{(2)}_{e\rm ph}{\check{\g}}_{\lambda},
\end{equation}
where $\check{\Si}^{(2)}_{e\rm ph}=\check{\Si}_{H}+\check{\Si}_{F}$ is the Hartree-Fock self-energy, depicted diagrammatically in Fig.~\ref{fig:diagrams:HF}, with
 \begin{equation}\label{eq:SigmaH}
\Si_{H}^{\eta \bar{\eta}}=\delta_{\eta \bar{\eta}}\sum_{\nu=\pm}\nu\eta \sum_{\ell}\M_{\ell}{\rm Tr}\{\n^{\nu}_{\bl}\M_{\ell} \}d^{\eta \nu}_{\ell}(\e=0)
\end{equation}
 \begin{equation}\label{eq:SigmaF}
\Si_{F}^{\eta \bar{\eta}}(\e)=i\sum_{\ell}
\int\frac{d\e'}{2\pi}\,d^{\eta \bar{\eta}}_{\ell}(\e-\e')\M_{\ell} \,
\g^{\eta \bar{\eta}}_{\bl}(\e')\M_{\ell},
\end{equation}
and $\eta, \bar{\eta}=\pm$. Above,  $d_{\ell}^{\eta
\bar{\eta}}(\e)$ represent the {\em free} thermalized phonon Green's
functions of the $\ell$-th phonon mode
\begin{equation}\nonumber
\label{eq:phononGF}
\begin{split}
 d_{\ell}^{\pm\pm}(\e) &=\sum_{s=\pm}\big[-i\pi (2\mathcal{N}_{\ell}+1)\delta(\e+s\hbar\om_{\ell })\pm \mathcal{P}\frac{s}{\e+s\hbar\om_{\ell }} \big]\\
 d_{\ell}^{\mp,\pm}(\e)&=-2\pi i [(\mathcal{N}_{\ell }+1)\delta(\e \pm \hbar\om_{\ell })+\mathcal{N}_{\ell }\delta(\e \mp \hbar \om_{\ell })]
\end{split}
\end{equation}
with $\mathcal{N}_{\ell }\equiv(e^{\beta\hbar \om_{\ell}}-1)^{-1}$
the thermal expectation value of the mode occupation. The proper
inclusion of possible heating effects on noise, involving non-equilibrium phonon occupation and its potential back-action on the electronic transport, is beyond the scope
of this publication; some of the involved issues are
discussed in the concluding Sec.~\ref{Sec:concl}.

In Eq.~\eqref{eq:SigmaH} we introduced the generalized electronic
density $\n_{\bl}^{\nu}$ on the two branches of the Keldysh
contour ($\nu=\pm$) in the presence of the counting field
\begin{equation}
\begin{split}
\n_{\bl}^{\nu}&\equiv \lim_{t\to t'+0^{\nu}} -i \hbar\, \g^{\nu \nu}_{\bl} (t-t')\\
     &=\lim_ {t\to t'+0^{\nu} }-i\int\frac{d\e}{2\pi}e^{-i \e  (t-t')/\hbar}\g^{\nu \nu}_{\bl}(\e).
\end{split}
\end{equation}
Note that, on the two branches of the Keldysh contour, the electronic density $\n_{\bl}^{\nu}$ 
is given by {\em different} limits $t\to t'$ of the corresponding Green's functions $\g^{\nu \nu}_{\bl} (t-t')$.
As a consequence, even if $\lambda(t)=\pm \lambda$ on the forward/backward branch, 
 $n^{+}_{\bl}\neq n^{-}_{-\bl}$
for any finite value of $\bl$.  On the other hand, at $\bl=0$ one gets
\begin{equation}\label{eq:density}
\n^{+}_{\bl=0}= \n^{-}_{\bl=0}=\n_{e}\equiv-i\int\frac{d\e}{2\pi}\g^{- +}_{\bl=0}(\e),
\end{equation}
where $\n_{ e}$ is the electronic density in the noninteracting
case.

\begin{figure}
\begin{center}
{\resizebox{0.95\columnwidth}{!}{\includegraphics{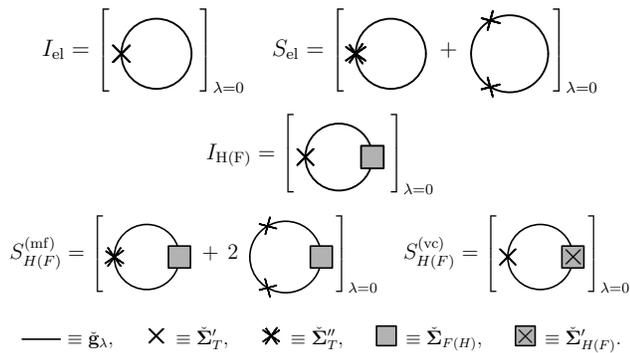}}}
\end{center}
\caption{Diagrammatic representations of Eqs.~\eqref{eq:I_el}-\eqref{eq:S_FH}. The plain line stands for the electronic Green's function $\check{\g}_{\bl}$; single and doubled crosses stand for $\check{\Si}_{T}'$ and $\check{\Si}_{T}''$, respectively. The box represents $\check{\Si}_{H(F)}$ and, finally, the crossed box stands for the derivative of the $e$-ph self-energy with respect to the counting field $\check{\Si}_{H(F)}'$.   }\label{fig:diagramsSmfvc}
\end{figure}

Plugging Eqs.~\eqref{eq:SigmaH}, \eqref{eq:SigmaF} into
Eq.~\eqref{eq:Gapprox}, we can rewrite Eqs.~\eqref{eq:I}, \eqref{eq:S} as $I= I_{\rm
el}+I_{e\rm ph}$ and $S= S_{\rm el}+S_{e\rm ph}$, where
\begin{equation}\label{eq:I_el}
I_{\rm el}=i e \int
\frac{d \e}{2 \pi \hbar} {\rm Tr}_{K}\left\{ \check{\Si}_{T}'\check{\g}_{\bl} \right\}_{\bl=0},
\end{equation}
\begin{equation}\label{eq:S_el}
S_{\rm el}= e^{2} \int
\frac{d \e}{2 \pi \hbar} {\rm Tr}_{K}\left\{ \check{\Si}_{T}''\check{\g}_{\bl}+\check{\Si}_{T}' \check{\g}_{\bl}\check{\Si}_{T}' \check{\g}_{\bl} \right\}_{\bl=0},
\end{equation}
are the elastic current and noise, and
\begin{equation}\nonumber
I_{e\rm ph}=I_{F}+I_{H}, \qquad \ S_{e\rm ph}=S_{F}+S_{H}
\end{equation}
give  the respective corrections due to electron-phonon coupling,  with
\begin{equation}\label{eq:I_FH}
I_{H(F)}=i e \int \frac{d \e}{2 \pi \hbar} {\rm Tr}_{K}\left\{
\check{\Si}_{T}'{\check{\g}}_{\lambda}\check{\Si}_{H(F)}{\check{\g}}_{\lambda} \right\}_{\bl=0},
\end{equation}
and $S_{H(F)}=S_{H(F)}^{\rm (mf)}+S_{H(F)}^{\rm(vc)}$, where
\begin{subequations}\label{eq:S_FH}
\begin{equation}
\begin{split}
S_{H(F)}^{\rm(mf)}= e^{2}\! \int\! \frac{d \e}{2 \pi \hbar} {\rm
Tr}_{K}\!\Big\{\,&\check{\Si}_{T}''{\check{\g}}_{\lambda}\check{\Si}_{H(F)}{\check{\g}}_{\lambda}\\
 +\,
2\,&\check{\Si}_{T}' \check{\g}_{\bl}\check{\Si}_{T}'
{\check{\g}}_{\lambda}\check{\Si}_{H(F)}{\check{\g}}_{\lambda} \!  \Big\}_{\bl=0}
\end{split}
\end{equation}
is the mean-field contribution and
\begin{equation} 
S_{H(F)}^{\rm(vc)}=e^{2} \int \frac{d \e}{2 \pi \hbar} {\rm
Tr}_{K}\left\{ \check{\Si}_{T}'
\check{\g}_{\bl}\check{\Si}_{H(F)}' \check{\g}_{\bl}
\right\}_{\bl=0}.
\end{equation}
\end{subequations}
the vertex correction. The previous equations can be schematically 
expressed by the diagrams of Fig.~\ref{fig:diagramsSmfvc}.

We note that truncating the Dyson equation to the lowest order in $e$-ph coupling, $\mathcal{O}(M^{2})$,  preserves charge conservation in that order. This implies that both mean current and zero-frequency noise are constant throughout the whole circuit, in particular at both junctions between the device and the leads. Potential violations of charge conservation can only occur in the next order, i.e. $\mathcal{O}(M^{4})$,  and can thus be safely neglected for any practical purposes in the considered weak coupling limit.  


Finally, we observe that to the lowest order in the $e$-ph coupling,
$I_{e\rm ph}$ and $S_{e\rm ph}$ are simply given by a linear
superposition of contributions coming from different phonon modes.
As a consequence, we  can restrict ourselves to the case of
coupling to a single phonon mode with frequency $\om_{0}$,
occupation $\np\equiv(e^{\beta\hbar  \om_{0}}-1)^{-1}$ and
coupling matrix $\M$.

\subsection{The extended wide band limit}
\label{Subsec:eWBL}
The corrections to current \eqref{eq:I_FH} and noise
\eqref{eq:S_FH} due to the $e$-ph coupling involve energy integrals
which can be evaluated in general only numerically. Analytical
progress can still  be  made if one assumes the electronic structure
to be slowly changing over few multiples of a typical phonon
energy around the Fermi level $E_F$ and approximate (i) the level
broadening $\Ga_{\alpha}$ and (ii) the non-interacting
retarded/advanced Green's function $\g^{r(a)}$ with their values
at the Fermi
energy~\cite{Paulsson:RapCom,Frederiksen:PRB07,Viljas,delaVega}
\begin{equation*}
\Ga_{\alpha}(\e)\approx\Ga_{\alpha}(E_F)\equiv\Ga_{\alpha}, \quad
\g^{r(a)}(\e)\approx\g^{r(a)}(E_F)\equiv\g^{r(a)},
\end{equation*}
where we took $\g^{r(a)}\equiv
\g^{--}_{\bl=0}-\g^{-+(+-)}_{\bl=0}$ as the definition of $\g^{r
(a)}$. This approximation, which we call ``extended wide-band
limit'' (eWBL), is reasonable for systems where either the
broadening due to tunneling is large ($\Gamma \gg  eV, k_B T$, and
$\hbar \om_{0}$), or  the closest resonance energy $\e_{\rm
res}$ is far away from the Fermi energy ($|\e_{\rm res}-E_{F}|\gg
\Gamma, eV, k_B T$, and $\hbar \om_{0}$).

Within the eWBL approximation, the integration over energy of
functions with compact support can be performed analytically,
resulting in explicit results for the mean current and the noise
as functions of the applied bias voltage and other system
parameters. It should be noted however, that approximation (ii) potentially leads 
to problems for integrals over infinite range and, in this case, 
it might be necessary to lift it.  Specifically, this happens in 
the calculation of the electron density entering
the Hartree term, see Appendix ~\ref{App:ne}, and in the evaluation of
the real parts of the retarded/advanced Fock self-energy via
Kramers-Kronig relations, see Appendix ~\ref{App:selfenergy}.


\section{Elastic current and noise}
\label{Sec:elastic}
For sake of completeness, before discussing the corrections to $I$  and $S$  due to the $e$-ph coupling, we consider briefly the
results for the {\em elastic} current and noise.

In the eWBL approximation, the elastic current is simply proportional to the voltage
\begin{equation} \label{eq:I_{el}_{T}}
I_{\rm el}=\frac{e}{h}{\rm Tr}\{\Tlr\}eV,
\end{equation}
with  $ \Tlr={\Ga_{L}\g^{r}\Ga_{R}\g^{a}}$, while the noise is given by
\begin{equation} \label{eq:S_{el}_{T}}
S_{\rm el}=\frac{e^{2}}{h}\left[ \frac{2}{\beta}{\rm Tr}\{\Tlr^{2}\}+{\rm Tr}\{\Tlr({\bf 1}-\Tlr)\}U(eV)\right],
\end{equation}
where we have introduced the function $U(x)=x\coth(\beta x/2)$.

The eigenvalues of the matrix $\Tlr$ give the ``PIN-code" of
transmission eigen-channels of the molecule connected to leads
(without $e$-ph interaction), and Eqs.~\eqref{eq:I_{el}_{T}},
\eqref{eq:S_{el}_{T}} are indeed equivalent~\cite{Meir} to the
standard results for current and noise in a non-interacting
system derived within the scattering
theory.~\cite{Buettiker,Blanter} However, $\Tlr$ is {\em not}
equal to the matrix product $\mathbf{t}\mathbf{t}^{\dagger}$ of the
transmission amplitudes $\mathbf{t}$ of the scattering theory (it
cannot be as $\Tlr$ is in general non-hermitian, for example). The
two matrices are related though by a similarity (non-unitary)
transformation, which among others ensures
$\Tr{\Tlr}=\Tr{\mathbf{t}\mathbf{t}^{\dagger}}$.~\cite{Meir} 
With this caveat in mind, for
sake of simplicity  we will
nevertheless call $\Tlr$ the transmission matrix in the rest of
this paper. The construction of the scattering eigenstates within
the NGF formalism is described in detail in
Ref.~\onlinecite{PaulssonBrandbyge}.\\

We now turn our attention to  the corrections to the current and noise
induced by the $e$-ph interaction. In order to make the discussion as
clear as possible, we will consider the contributions coming from
the Hartree and the Fock diagrams separately.


\section{Corrections due to the Hartree diagram}
\label{Sec:Hartree}
\subsection{Current}
We start by considering the contributions to the current  $I_{H}$
coming from the Hartree diagram. After integrating Eq.~\eqref{eq:I_FH}
in the eWBL approximation we obtain
\begin{equation}\label{eq:IH}
I_{H}=\frac{e}{h}{\rm Tr}\{\Tlr_{H}^{\rm (qel)} \}eV,
\end{equation}
with
\begin{equation}
\Tlr_{H}^{\rm (qel)}=-\frac{2 {\rm Tr}\{ \n_{e}  \M \}}{\hbar
\om_{0}}\, \Ga_{L}(\g^{r} \M \A_{R}+h.c.)
\end{equation}
with $\n_{e}$ the noninteracting electron density [
Eq.~\eqref{eq:density}] and $\A_{\alpha}=\g^{r}\Ga_{\alpha}\g^{a}$. The
correction $I_{H}$ is therefore a smooth function of the voltage
with no features at the phonon emission threshold. For this reason
$I_{H}$ has been often discarded in previous works on the effects of $e$-ph
interaction on the current.~\cite{Frederiksen:PRL04,Paulsson:RapCom,Frederiksen:PRB07,Paulsson:PRL100,Viljas,delaVega,
Galperin:PRB06,Fransson} 

It should be noticed, however, that
$I_{H}$ is generally non-linear in $eV$, since $\n_{e}$ can be a
(smooth) function of the applied bias voltage. Such a voltage
dependence is nevertheless rather weak in the eWBL (see
Appendix ~\ref{App:ne}), and in such a case
it is possible to interpret Eq.~\eqref{eq:IH} as a quasi-elastic
correction to an effective transmission matrix
$\Te=\Tlr+\Tlr_{H}^{\rm (qel)}$, i.e.\ because of the $e$-ph
coupling, the current is not proportional to the bare transmission
coefficient ${\rm Tr}\{\Tlr \}$ but rather to ${\rm Tr}\{\Te \}$.


\subsection{Noise}

\begin{figure}
\begin{center}
{\resizebox{0.98\columnwidth}{!}{\includegraphics{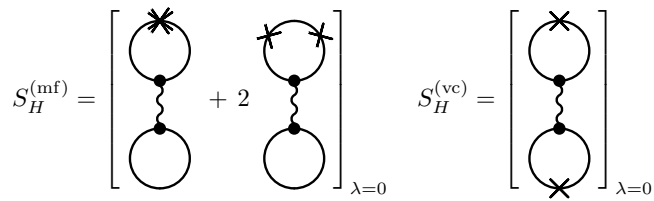}}}
\end{center}
\caption{Diagrammatic representations of $S_{H}^{\rm (mf)}$ and $S_{H}^{\rm (vc)}$. Here, plain and wiggly lines stand for the electronic  and phononic Green's functions, respectively. The single cross stands for $\check{\Si}_{T}'$ and the doubled one for $\check{\Si}_{T}''$. Finally, the dot represents the $e$-ph coupling constant $\M$. }\label{fig:diagramsSH}
\end{figure}

The mean-field contributions and the vertex correction to noise
due to the Hartree diagram can be schematically represented by the
diagrams in Fig.~\ref{fig:diagramsSH}, which are the result of inserting the Hartree self-energy from Fig.~\ref{fig:diagrams:HF} into appropriate diagrams in Fig.~\ref{fig:diagramsSmfvc}. In the usual eWBL,
$S_{H}^{\rm(mf)}$ takes the simple form
\begin{equation} \label{eq:SHmf}
\begin{split}
\frac{S_{H}^{\rm (mf)}}{e^{2}/h}=\Tr{(\mathbf{1}-2\Tlr)\Tlr_{H}^{\rm (qel)}} U(eV)+\frac{4}{\beta}\Tr{\Tlr\Tlr_{H}^{\rm (qel)}}.
\end{split}
\end{equation}
Analogously to the current $I_{H}$, this contribution has a simple
interpretation in terms of the renormalization of the transmission
matrix introduced above $\Tlr \to \Te=\Tlr+\Tlr_{H}^{\rm (qel)}$.
This can be seen easily, as  Eq.~\eqref{eq:SHmf} corresponds
exactly to the contribution of order $\M^{2}$ to the elastic
shot-noise of a system with transmission matrix $\tilde{\Tlr}$:
\begin{equation*}
\begin{split}
&\frac{2}{\beta}\Tr{\Te^{2}}+\Tr{\Te({\bf 1}-\Te)}U(eV)=\frac{h}{e^{2}}S_{\rm el}+\\
&+\Tr{(\mathbf{1}-2\Tlr)\Tlr_{H}^{\rm (qel)}} U(eV)+\frac{4}{\beta}\Tr{\Tlr\Tlr_{H}^{\rm (qel)}}+\mathcal{O}(\M^{4}),
\end{split}
\end{equation*}
where $S_{\rm el}$ is given in Eq.~\eqref{eq:S_{el}_{T}}.

Making use of the cyclic invariance of the trace, the vertex correction  $S_{H}^{\rm
(vc)}$  can be rewritten as
\begin{equation}
\begin{split}\label{eq:SHvc}
\frac{S_{H}^{\rm (vc)}}{e^{2}/h}&=\frac{2 i}{\hbar \om_{0}} \left[\left({\rm Tr}\left\{\M \n_{+}'\right\}\right)^{2}-\left({\rm Tr}\left\{\M \n_{-}'\right\}\right)^{2} \right]\\
&=\frac{8}{\hbar \om_{0}}  {\rm Re} \left[ \Tr{ \M \n_{-}'  } \right]  {\rm Im} \left[ \Tr{ \M \n_{-} '} \right],
\end{split}
\end{equation}
where we have used the fact that
$\left[\n_{-}'\right]^{\dag}=-\n_{+}'$, with
$\n_{\nu}'\equiv(\partial \n_{\bl}^{\nu}/\partial \bl)_{\bl=0}$.
Performing the integrals over energy in the usual eWBL
approximation\footnote{Unlike $ \n_{e}$, the evaluation of
$\n_{\nu}'$ involves only integrands with compact support.} one
obtains
\begin{align*}
 {\rm Re}\left[\Tr{\M \n_{-}'}\right]&=-\frac{1}{2}\Tr{\Ga_{L}\A_{R}\M\g^{a}+h.c.}eV,\\
 {\rm Im}\left[\Tr{\M \n_{-}'}\right]&=-\frac{i}{2}\Tr{\Ga_{L}\g^{r}_{R}\M\A_{L}-h.c.}\frac{2}{\beta} \\
 &-\frac{i}{2}\Tr{\Ga_{L}\g^{r}_{R}\M\A_{R}-h.c.}U(eV)\\
 & +\frac{1}{4}\Tr{\Ga_{L}(\A_{L}\M\A_{R}\!-\!\A_{R}\M\A_{R}\!+\!h.c.\!)}\\
 &\times \left(\!\frac{2}{\beta}\!-\! U(eV)\!\!\right) \nonumber
\end{align*}
with $\g_{R}^{r}={\rm Re}\,\g^{r}$. Contrary to $S_{H}^{\rm (mf)}$, Eq.~\eqref{eq:SHvc}
has no simple interpretation in terms of an effective transmission
coefficient and it represents a distinctive contribution to noise
coming from the Hartree term. 
From the physical point of view, it stems from the 
coupling of occupations of the electronic levels with the current
fluctuations.\cite{GogolinKomnik,Hershfield} 

We note, however, that ${\rm Im}[\Tr{\M \n_{-}'}]=0$ in the case of a system with a
single electronic level symmetrically coupled to leads. Therefore
in this particular case the correction to noise induced by the
Hartree term is given by $S_{H}^{\rm(mf)}$ alone.


\section{Corrections due to the Fock diagram}
\label{Sec:Fock}
\subsection{Current }
We now turn our attention to the corrections to current induced by the Fock diagram.
Integrating Eq.~\eqref{eq:I_FH} in the usual eWBL approximation, we obtain
\begin{equation}\label{eq:I_{F}}
\begin{split}
\frac{I_{\rm F}}{e/h}&= {\rm Tr}\{\Tlr_{F}^{(\rm qel)} \}eV +{\rm Tr}\{ \Tlr_{F}^{\rm (inel)}\} g(eV)\\
&+2\np{\rm Tr}\{\Tlr_{F}^{(\rm qel)}+\Tlr_{F}^{(\rm inel)} \} eV +
{\rm Tr}\{ \Tlr_{F}^{\rm (asym)}\}h(eV)
\end{split}
\end{equation}
where
\begin{subequations}
\begin{align}
&\Tlr_{F}^{(\rm qel)} \!=\Ga_{L}(\g^{r}\M \g^r_{R}\M\A_{R}+h.c.) \label{eq:TFqel}\\
&\Tlr_{F}^{(\rm inel)}\! =\Ga_L\g^r\Big[\M \A_R \M\! -\!\frac{i}{2}(\M\A\M\g^r\Ga_R\!-\!h.c)\! \Big]\g^a \label{eq:TFinel} \\
&\Tlr_{F}^{\rm (asym)}\!=\Ga_L\g^r\big[\M
(\A_L-\A_R)\M\g^r\Ga_R+h.c.\big]\g^a
\end{align}
\end{subequations}
with $\A\equiv\A_{L}+\A_{R}=i(\g^{r}-\g^{a})$ the spectral density. All the involved quantities depend only on the properties of the
system at the Fermi level and can be determined by {\em
ab-initio} calculations.~\cite{Paulsson:RapCom,Frederiksen:PRB07,Viljas,delaVega}

The voltage dependence of $I_{F}$ is carried by the functions
\begin{equation}\label{eq:g}
g(eV)= \frac{1}{2}\big[U(eV\!-\! \hbar\om_{0})\!-\!U(eV\!+\!\hbar\om_{0})\!+\!2eV\big],
\end{equation}
and
\begin{equation}\label{eq:h}
\begin{split}
h(eV)&=\frac{1}{2}\int{d\e}\big[\big(n_F(\e)-n_F(\e+eV)\big)\\
&\times \Ht\{n_F(\e' -\hbar\om_0)-n_F(\e' +\hbar\om_0) \}(\e)\big],
\end{split}
\end{equation}
where $\Ht\{ f(\e')\}(\e)=(1/\pi)\, \mathcal{P}\int d\e' f(\e')/(\e'-\e)$ is the Hilbert transform.
Eq.~\eqref{eq:I_{F}} is in 
agreement with the result of Viljas {\em et al.},~\cite{Viljas} while a term  $\propto
(1+2\np){\rm Tr}\{\Tlr_{F}^{(\rm qel)} \}eV $ is missing in Refs.~\onlinecite{Paulsson:RapCom, Frederiksen:PRB07}.
Such a discrepancy is further discussed in Appendix ~\ref{App:selfenergy}. 

\begin{figure}
\begin{center}
{\resizebox{0.85\columnwidth}{!}{\includegraphics{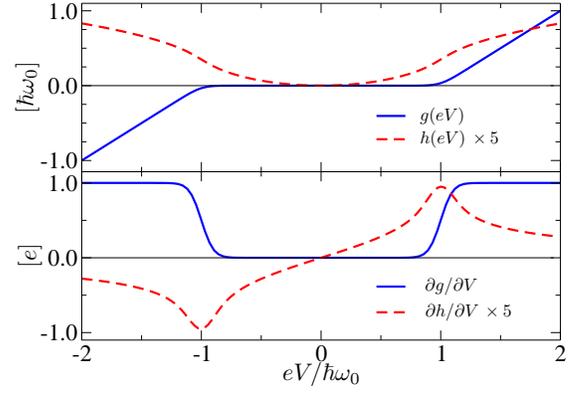}}}
\end{center}
\caption{Upper panel: Plots of the dependencies of the functions $g(eV)$ and $h(eV)$ on the applied bias voltage (Eqs.~\eqref{eq:g} and \eqref{eq:h}, respectively). 
Lower panel: Same as above, but for the derivatives $\partial g/\partial V$ and $\partial h/\partial V$. In both panels  $k_{B}T=\hbar\om_{0}/30$. 
}\label{fig:plot:gh}   
\end{figure}

The functions $g(eV)$ and $h(eV)$ give contributions to $dI/dV$ which are 
even/odd in bias, respectively (see Fig.~\ref{fig:plot:gh}). 
The term proportional to $h(eV)$ vanishes  in the case of symmetric
coupling to the leads, and it is typically much smaller than the
contribution proportional to $g(eV)$, even for asymmetric
junctions.~\cite{Frederiksen:PRB07,Paulsson:PRL100} Moreover,
experimentally measured conductance curves are usually very weakly
asymmetric under reversal of $V$ and at present it is unclear if
the asymmetry is caused by phonons or by other effects.

At low temperature ($k_{B}T\ll\hbar \om_{0}$, $\np\approx 0$), the
main contribution to $I_{F}$ is therefore given by the first two
terms of Eq.~\eqref{eq:I_{F}} alone. The first of these terms,  linear in $eV$, is a quasi-elastic correction that, similarly
to $I_{H}$, contributes to an effective transmission matrix
$\Te=\Tlr+\Tlr_{F}^{\rm (qel)}$. The second one has instead a
threshold behavior at the phonon emission energy, see Fig.~\ref
{fig:plot:gh}, and it is responsible for the jump in the
conductance observed in IETS and PCS experiments.

The sign of the conductance step at the phonon emission threshold
(positive or negative) depends on the coefficient ${\rm
Tr}\{\Tlr_{F}^{\rm (inel)}\}$, and it has been discussed in detail
in Refs.~\onlinecite{Paulsson:RapCom, Viljas,delaVega}.  As a rule of thumb,
in the case of a molecular junction with low (high) bare
transmission ${\rm Tr}\{\Tlr\}$,  inelastic $e$-ph scattering
results in an increase (decrease) of the conductance above the
phonon emission threshold.

In the case of a system with a single electronic level
symmetrically coupled to the leads via $\Gamma_{L}=\Gamma_{R}\equiv\Gamma$, $\Tlr_{F}^{\rm (inel)}$
reduces to $\Tlr_{F}^{\rm
(inel)}=(M^{2}\T^{2}/\Gamma^{2})(1-2\T)$, where
$\T=\Gamma^{2}|g^{r}|^{2}$ is the transmission coefficient.   In
this case, the crossover from an increase to a decrease in the
conductance is predicted to occur at
$\T=1/2$.~\cite{Paulsson:RapCom,delaVega} This behavior has been
explored and confirmed experimentally in Ref.~\onlinecite{Tal}.

\subsection{Noise} \label{sec:SF}
\begin{figure}
\begin{center}
{\resizebox{0.98\columnwidth}{!}{\includegraphics{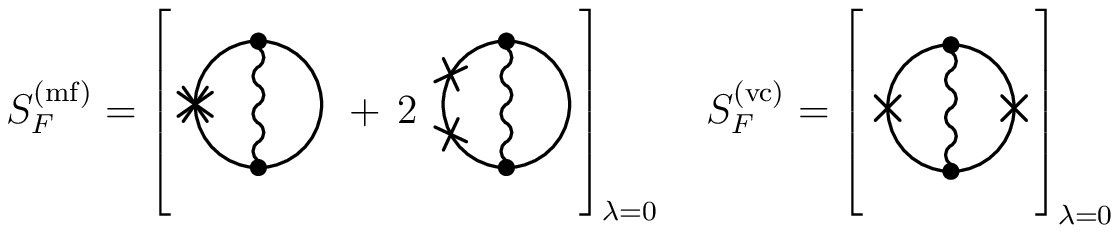}}}
\end{center}
\caption{Diagrammatic representations of $S_{F}^{\rm (mf)}$ and $S_{F}^{\rm (vc)}$. As in Fig.~\ref {fig:diagramsSH},  plain and wiggly lines stand for the electronic and phononic Green's functions, respectively. The single cross stands for $\check{\Si}_{T}'$ and the doubled one for $\check{\Si}_{T}''$. Finally, the dot represents the $e$-ph coupling constant $\M$.  }\label{fig:diagramsSF}
\end{figure}

We finally address the corrections to noise due to the Fock diagram
$S_{F}=S_{F}^{\rm (mf)}+S_{F}^{\rm (vc)}$, which are schematically represented by the diagrams in Fig.~\ref{fig:diagramsSF}.

After lengthy but straightforward calculations, integration over
energy in the usual eWBL approximation leads to analytic results
for $S_{F}^{\rm (mf/vc)}$ as functions of the applied bias
voltage. The final expressions are, however, rather cumbersome and, for
simplicity, we consider here only the limit of zero
temperature $T=0$. The complete expressions for $S_{F}^{\rm
(mf/vc)}$ at finite temperature are given in the supplementary material,~\footnote{See associated \texttt{Mathematica} notebook for the complete expression for $S_{F}^{\rm (mf/vc)}$ at finite temperature.} while the limit $eV\to 0$ is discussed in Appendix ~\ref{App:FDT} in relation 
to the fluctuation-dissipation theorem.

In the limit of zero temperature, we obtain
\begin{align}\label{eq:SFmf}
\frac{S_{F}^{\rm (mf)}}{e^{2}/h}&=\Tr{(\mathbf{1}-2\Tlr)\Tlr_{F}^{\rm (qel)}} |eV|  \nonumber\\
                          &+\Tr{(\mathbf{1}-2\Tlr)\Tlr_{F}^{\rm (inel)}}(|eV|-\hbar\om_{0})\theta (|eV|-\hbar\om_{0}) \nonumber \\
                          &+\Tr{\K_{1}^{\rm (mf)}}\mathrm{sign}(eV) h(eV)\big|_{T=0},
\end{align}
and
\begin{equation}\label{eq:SFvc}
\begin{split}
\frac{S_{F}^{\rm (vc)}}{e^{2}/h}&=\Tr{\Q_{F}^{\rm (inel)}}(|eV|-\hbar\om_{0})\theta (|eV|-\hbar\om_{0})\\
        &+\Tr{\K_{1}^{\rm (vc)}}\mathrm{sign}(eV) h(eV)\big|_{T=0},
\end{split}
\end{equation}
where $\Tlr_{F}^{\rm (qel/inel)}$  are given in Eqs.~\eqref{eq:TFqel}, \eqref{eq:TFinel}, 
\begin{equation}
\begin{split}
\Q_{F}^{\rm (inel)}=&-\g^{a }\Ga_{L}\g^{r}\big[ \M \A_{R}\Ga_{L}\A_{R}\M\\&+ \M \A_{R}\Ga_{L}\g^{r}\M\g^{r}\Ga_{R}
+h.c.\big].
\end{split}
\end{equation}
and 
\begin{subequations} \label{eq:K1}
\begin{align}
&\K_{1}^{(\rm mf)}\!=\!(\mathbf{1}\!-2\Tlr)\Ga_{L}\!\big[\A_{R}\M(\A_L\!-\!\A_{R})\M\g^{a}\!+\!h.c.\big]\\
&\K_{1}^{(\rm vc)}\!=\M(\A_{R}\Ga_{L}\g^{r}\!+\g^{a}\Ga_{L}\A_{R})\M\\
&\phantom{\K_{1}^{(\rm vc)}}\times\big[\A_{R}\Ga_{L}(\A_{L}-\A_{R}\!+2i\g_{R}^{r})\!+h.c.\big]. \nonumber
\end{align}
\end{subequations}
Finally, 
\begin{equation}
\begin{split}
h(eV)\big|_{T=0}=&\frac{\hbar \om_{0}}{2}\sum_{s=\pm 1}s\left (\frac{eV}{\hbar\om_{0}}+s \right) \ln \left| \frac{eV}{\hbar\om_{0}}+s \right|
\end{split}
\end{equation}
is the zero temperature limit  of Eq.~\eqref{eq:h}. 
The corrections to noise  $S_{F}^{\rm (mf/vc)}$ can then be divided 
into a symmetric term, which is even in bias, and an antisymmetric one, 
which contains the Hilbert transform $h(eV)\big|_{T=0}$ and yields an odd contribution. 
We notice that, while $h(eV)\big|_{T=0}$ is a continuous function, its derivative shows logarithmic divergencies at $eV=\pm \hbar \om_{0}$.\cite{Entin-Wohlman,Egger}
These zero-temperature divergencies are, however, an artifact of treating  the phonons 
as non-interacting modes, and they are regularized either by finite temperature or if 
any broadening of the phonon spectrum is taken into account.\cite{Egger,Entin-Wohlman} This issue, however, goes beyond the scope of this work.   

At  zero temperature,  the symmetric contribution to $S_{F}$ is 
a piece-wise linear function of $eV$. At low voltages, $|eV|< \hbar\om_{0}$, it is given
by the first term of Eq.~\eqref{eq:SFmf} alone. Following the same
reasoning as for  Eq.~\eqref{eq:SHmf}, this linear contribution
can be directly interpreted in terms of the renormalization of the
transmission $\Tlr\to\tilde{\Tlr}=\Tlr+\Tlr^{\rm(qel)}_{F}$, 
consistently with the sub-threshold correction to the current. 
Above the phonon emission threshold, $|eV|>\hbar\om_{0}$, inelastic
processes come into play and their contribution to the noise is
given {\em both} by the second term of Eq.~\eqref{eq:SFmf} and by
the vertex correction Eq.~\eqref{eq:SFvc}. It is important to
notice that these two contributions are in general of the same
order (see below the Sec. \ref{Subsec:indeplev} for a demonstrative example), so that the latter cannot be discarded.

Experimentally,  $\tfrac{\partial S}{\partial V}$ is often measured directly by a lock-in technique. Such a quantity shows at the phonon emission threshold
a sharp and distinguishable jump on top of  a featureless background due to the  elastic and quasi-elastic contributions.
Therefore, we define here the {\em inelastic noise signal} as the difference of the plateau values of the noise derivative just above and below the jump
\begin{equation}
\begin{split}
\Delta S'=&\frac{\partial S}{\partial V}\Big|_{|eV|=\hbar \om_{0}+c k_{B}T}-\frac{\partial S}{\partial V}\Big|_{|eV|=\hbar \om_{0}-c k_{B}T}
\end{split}
\end{equation}
with $c\sim 5$ accounting for the finite jump width at finite temperatures. At low enough temperatures, 
terms proportional to $h(eV)$ give a very small contribution to the inelastic noise signal due to the symmetric shape of $\partial_{V}h$ around $|eV|=\hbar\om_{0}$ (for details see Appendix ~\ref{App:asym})
and we can then approximate
\begin{equation}
\Delta S'\approx \frac{e^{3}}{h}\Tr{(\mathbf{1}-2\Tlr)\Tlr_{F}^{\rm
(inel)}+\Q_{F}^{\rm (inel)}},
\end{equation}
i.e.\ at low temperatures $\Delta S'$ carries the structural information about
the junction given by the terms with the threshold
behavior at the phonon emission energy.

\subsection{Independent electronic levels}
\label{Subsec:indeplev}

\begin{figure}
\begin{center}
{\resizebox{0.85\columnwidth}{!}{\includegraphics{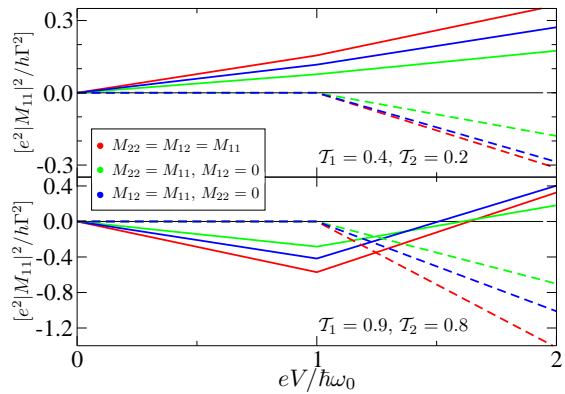}}}
\end{center}
\caption{(Color online) Voltage dependence of $S_{F}^{\rm (mf)}$ (full lines) and $S_{F}^{\rm (vc)}$ (dashed lines) for the case of the toy model (Eqs.~\eqref{eq:SFTmf} and \eqref{eq:SFTvc}, respectively) for different values of the transmission coefficients and of the $e$-ph coupling matrix elements. Upper panel: $\T_{1}=0.4, \, \T_{2}=0.2$, lower panel $\T_{1}=0.9, \, \T_{2}=0.8$. In both panels $k_{B}T=0,$ and $\Gamma_{1}=\Gamma_{2}=\Gamma$. }
\label{fig:SvsV}
\end{figure}

We now consider a toy model for molecular junctions, in which we
assume the electronic levels to be mutually coupled only via the $e$-ph
interaction. In this case, the relevant matrices in the system electronic
space are given by
\begin{align*}
[\Ga_{L(R)}]_{ij}=\delta_{ij}\Gamma_{i,L(R)}, \quad [\g^{r}]_{ij}=\frac{\delta_{ij}}{\Delta_{i}+i(\Gamma_{i,L}+\Gamma_{i,R})/2},
\end{align*}
 and $[\M]_{ij}=M_{ij}$, where $i,j=1,\dots, N$, and 
 $N$ is the number of electronic levels involved in the transport.  
Under the further assumption that each channel is symmetrically
coupled to the leads ($\Ga_{L}=\Ga_{R}=\Ga$),
the prefactors
$\Tr{\K_{1}^{\rm (mf/vc)}}$ vanish identically and
Eqs.~\eqref{eq:SFmf},~\eqref{eq:SFvc} can be rewritten in a
particularly suggestive form in terms of the transmission
probabilities $\T_{i}=\Gamma_{i}^{2}/(\Delta^{2}_{i}+\Gamma_{i}^{2})$ of the individual levels
\begin{subequations} \label{eq:SFT}
\begin{widetext}
\begin{equation} \label{eq:SFTmf}
\begin{split} 
\frac{S_{F}^{\rm (mf)}}{e^{2}/h}&=2 |eV| \sum_{i=1}^{N}\Big\{\gamma_{ii} (1-\T_i)(1-2\T_i)+\sum_{j> i}\gamma_{ij} [\T_i(1- 2\T_i)+\T_j (1-2 \T_j)]\sqrt{\frac{(1-\T_i) }{\T_i}\frac{(1-\T_j) }{\T_j}} \Big\}\\
            &+(|eV|-\hbar\om_{0})\theta(|eV|-\hbar\om_{0})\sum_{i=1}^{N}\Big\{\gamma_{ii}(1-2 \T_i)^2+ 2\sum_{j> i}\gamma_{ij}(1-2 \T_i(1-\T_i)-2 \T_j(1-\T_j))\Big\},\\
\end{split}
\end{equation}
\begin{equation} \label{eq:SFTvc}
\begin{split}
\frac{S_{F}^{\rm (vc)}}{e^{2}/h}&=\!-2(|eV|-\hbar\om_{0})\theta(|eV|-\hbar\om_{0})\sum_{i=1}^{N}\!\Big\{ \!2\gamma_{ii}\T_i(1-\! \T_i) +\!\sum_{j> i}\!\gamma_{ij} \Big[ \T_i\!+\!\T_j\!-2\T_i\T_j \!+ 2\sqrt{\T_i(1-\!\T_i) \T_j(1-\!\T_j) }\Big]  \! \Big\},
\end{split}
\end{equation}
\end{widetext}
\end{subequations}
where we have introduced the dimensionless coupling constants
$\gamma_{ij}={\big| M_{ij}\big|^2 \T_i \T_j}/({\Gamma_i
\Gamma_j})$. 
For $N=1$, Eqs.~\eqref{eq:SFT} reduce
directly to the result of Refs.~\onlinecite{Haupt, Schmidt, Avriller}.
The voltage dependence of $S_{F}^{\rm (mf)}$ and $S_{F}^{\rm (vc)}$ is presented 
in Fig.~\ref{fig:SvsV} for the case of a systems with only two levels. 
We notice that $S_{F}^{\rm (vc)}<0$ (see also Eq.\ref{eq:SFTvc}), meaning that the vertex corrections
correspond to processes that lead to a suppression of the
noise through the system. Moreover, Fig.~\ref{fig:SvsV} evidences that the contributions 
to the noise due to the vertex corrections can be of the same order of magnitude as the 
mean-field ones, and that they generally need 
to be taken into account in order to make accurate predictions for the phonon-assisted 
current noise.

\begin{figure}
\begin{center}
{\resizebox{0.98\columnwidth}{!}{\includegraphics{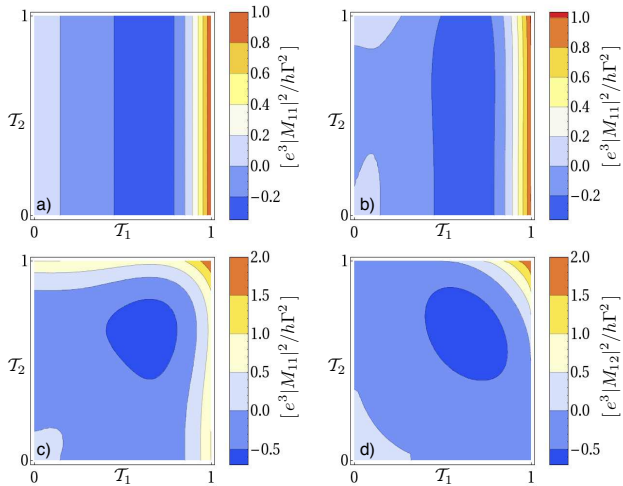}}}
\end{center}
\caption{(Color online) Contour plots of $\Delta S'$  at zero temperature as a function
of $\T_{1}$ and $\T_{2}$ for different values of the $e$-ph
coupling matrix elements $M_{ij}$. a) $M_{11}\neq 0$ and
$M_{22}=M_{12}=0$; b) $M_{11}\neq0$, $M_{22}=0.1M_{11}$ and
$M_{12}=0$; c)  $M_{11}\neq0$, $M_{22}=M_{11}$ and
$M_{12}=0$ ; d) $M_{12}\neq0$ and $M_{11}=M_{22}=0$. In all panels, $\Gamma_{1}=\Gamma_{2}=\Gamma$.}
\label{fig:DeltaS}
\end{figure}

In terms of the transmission coefficients of the different channels, the inelastic noise signal $\Delta S'$ is given by
\begin{equation}
\begin{split}\label{eq:DeltaS}
\Delta S'=&\frac{e^{3}}{h}\sum_{i=1}^{N}\big\{\gamma_{ii}(1-8 \T_i+8\T_{i}^{2})+\sum_{j> i}\gamma_{ij}\varphi(\T_{i},\T_{j})\big\}.
\end{split}
\end{equation}
with $\varphi(\T_{i},\T_{j})=2\big[(1-\T_{i}-\T_{j})^{2}- \T_i(1-\T_i) -
\T_j(1-\T_j)-2 \sqrt{\T_i\T_j(1-\T_i)(1-\T_j)}\,\big]$. 
Depending on the  values of $\T_{i}$ and $\gamma_{ij}$, $\Delta S'$ can be either
positive or negative and it is in general very sensitive to the
parameters  of the junction as illustrated in Fig.~\ref{fig:DeltaS}, again for the case 
of a system with only two levels. Here  we
plot $\Delta S'$ as a function of the transmission coefficients
$\T_{1}, \T_{2}$ for different values of the $e$-ph coupling
matrix elements. As general features we notice that $\Delta S'$ is always positive 
when $\T_{1},\T_{2}\ll1$ or when they are both close to the full transmission. Vice versa,
$\Delta S'$ is always negative and close to maximum suppression for $\T_{1},\T_{2}\approx0.5$. 
Interestingly, the characteristics of $\Delta S'$ depend strongly 
on the relative strength of the different matrix elements $M_{ij}$,
and therefore the inelastic noise signal might provide a tool to extract important 
information on the $e$-ph coupling. 

\section{Conclusions and Outlook}
\label{Sec:concl}

In conclusion, in this work we have studied the corrections 
due to weak electron-phonon coupling to the average current and 
the zero-frequency noise in a molecular junction. To address both 
quantities in a compact and efficient way,  we employed the 
generalized Keldysh Green's functions technique. Importantly, for 
the noise we were able to identify distinct terms 
representing the mean-field contribution and the vertex corrections, respectively. 
We included in our calculations both the contributions due to the Hartree
and to the Fock diagrams and, under the assumption that 
the densities of states of the system and of the leads depend weakly 
on energy (eWBL), we derived analytic expressions for $I_{H(F)}$ and $S_{H(F)}$
as functions of the applied bias voltage at arbitrary temperature. 
These expressions can serve as a basis for  {\em ab-initio} calculations 
to make realistic predictions for  the current noise in an experimentally significant 
class of molecular junctions.  Finally, we considered a toy model for molecular junctions
to elucidate the sensitivity of the inelastic phonon signal to the parameters characterizing the junction.

Throughout this paper we have assumed the phonon mode to be at
equilibrium with an external thermal bath, i.e. we have taken the
occupation $\np$ to be fixed according to the Bose-Einstein
distribution $\np=(e^{\beta\hbar\om_{0}}-1)^{-1}$. Such an
approximation is strictly consistent with the lowest-order
perturbation theory in the $e$-ph coupling when we implicitly
assume strong thermalization of the phonon mode. However, it turns out 
in practice that often heating effects cannot be  disregarded
and that they influence in turn the non-linear conductance.
\cite{Frederiksen:PRL04,Mitra,Viljas,Frederiksen:PRB07} 

From the theoretical point of view, the problem of non-equilibrium 
phonon heating can be addressed by extending the system Hamiltonian to 
include the coupling of the molecular phonon to other degrees of freedom 
(typically bulk phonons in the leads). The value of the 
corresponding coupling constants can ultimately  be obtained 
from  {\em ab-initio} calculations, which allow to asses the influence of the environment 
from a microscopical description.\cite{Engelund} 

In the case of {\em  zero} counting field $\bl=0$, the non-equilibrium phonon occupation
for weak $e$-ph coupling can be equivalently obtained either by a
full non-equilibrium calculation evaluating the phonon Green's function~\cite{Viljas,Asai,Mitra,Ryndyk} or by solving a master
equation describing the heating of the
device,~\cite{Frederiksen:PRL04,Mitra,Frederiksen:PRB07} which can be viewed as a kinetic-equation-like 
approximation to the full non-equilibrium Green's functions studies. 
Knowing the non-equilibrium phonon occupation allows to take consistently into account 
the effects of phonon heating in the non-linear conductance. In our pilot
study~\cite{Haupt} we used such an ingredient also to phenomenologically include heating
effects in the noise through a single level.~\footnote{Its generalization to
multilevel case is straightforward and follows exactly the lines of
Refs.~\onlinecite{Frederiksen:PRL04,Frederiksen:PRB07}.}

\begin{figure}
\begin{center}
{\resizebox{0.35\columnwidth}{!}{\includegraphics{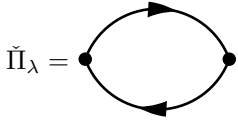}}}
\end{center}
\caption{Diagram corresponding to the polarization bubble for the
phonon $\check{\Pi}_{\bl}$. The plain lines and the dots represent the electronic Green's functions $\check{\g}_{\bl}$ and $e$-ph coupling constant $\M$, respectively.  }\label{fig:bubble}
\end{figure}

However, for a fully microscopical calculation of the noise, the situation is considerably more complicated because 
at {\em finite} counting field  $\bl\neq 0$, heating effects {\em cannot} 
be expressed solely in terms of the non-equilibrium occupation of the phonon mode. In fact,
to include phonon-heating effects in the generalized Keldysh GF technique
one has to solve the  Dyson equation for the phonon Keldysh Green's function
$\check{D}_{\bl}=\check{d}+\check{d}\,\check{\Pi}_{\lambda}\check{D}_{\bl}$,
with the polarization operator $\check{\Pi}_{\lambda}$ 
being given in the lowest order  by the electron-hole
bubble,~\cite{Viljas,Mitra,Urban} see Fig.~\ref{fig:bubble}.
Note that $\check{\Pi}_{\bl}$  
is explicitly $\bl$-dependent via
the electronic Green's functions and so is consequently also the dressed phonon
Green's function $\check{D}_{\bl}$.  At $\bl\neq 0$ the four Keldysh components of 
 $\check{D}_{\bl}$ are all independent and therefore, even in the kinetic limit (phonon
line-width neglected), it is not possible to express the effect of heating
just in terms of  a single non-equilibrium occupation.  It is important to notice that substituting
$\check{D}_{\bl}$  for the free phonon Green's function in the expressions for 
the $e$-ph self-energies $\check{\Si}_{H(F)}$ 
generates extra (additive) contributions to 
the vertex corrections $S_{H}^{\rm (vc)}$ and $S_{F}^{\rm (vc)}$. 
These contributions, which are related to the influence  of phonon
fluctuations on the electronic transport (``feedback"), are not included in our previous phenomenological 
treatment of heating effects on noise~\cite{Haupt} and they could possibly account for
the discrepancy between our result and an unpublished one by
Jouravlev and coworkers,~\cite{Jouravlev} which predicts the noise to grow with voltage  above the phonon 
emission threshold
as $S_{e\rm ph}^{(\rm Ref.\ 65)}\sim V^{4}$, 
in contrast to the quadratic behavior of Ref.~\onlinecite{Haupt}, $S_{e\rm ph}^{(\rm Ref.\ 46)}\sim V^{2}$. 
The idea that phonon heating effects could be responsible for a  nonlinear voltage dependence 
of  $\partial S_{e\rm ph}/\partial V$ is further corroborated by a recent work by Urban {\em et al.},~\cite{Urban} which, 
however, predicts $S_{e\rm ph}^{(\rm Ref. \ 64)}\sim V^{3}$.  An independent 
calculation is therefore required to settle this issue.  Careful inclusion of phonon heating effects 
into the noise calculations then certainly represents a relevant extension of our studies, furthermore urged by
the relevance of  heating in several  IETS and PCS experiments.

Very recently, a lot of interest has been paid to the study of current-induced excitations of  local spin degrees of freedom 
in spin-dependent IETS set-ups.~\cite{Eigler,Heinrich,Cinane,Fernandez-Rossier:PRL09, *Persson:PRL09, *Fransson:NL09, Lorente:PRL09, *Gauyacq:PRB10, *Novaes:preprint, Delgado:PRL10, *Delgado:preprint, Balatsky, Timm} 
Several of these calculations~\cite{Fernandez-Rossier:PRL09, *Persson:PRL09, *Fransson:NL09, Delgado:PRL10, *Delgado:preprint,Balatsky} rely on a perturbative 
approach analogous to the lowest order expansion of Ref.~\onlinecite{Paulsson:RapCom}, 
also used in this paper.  So far, those studies have dealt exclusively with the 
non-linear conductance and the study  of current noise in those 
spin systems would be a most natural next step. Our method can be
straightforwardly extended in this direction, as long as the
occupation of spin states is described in a phenomenological way
via the master equation\cite{Delgado:PRL10, *Delgado:preprint} (or just by thermal
distribution). However, possible further extensions to account for
fluctuations of a non-equilibrated spin remain, even conceptually, an
open question, because of the anharmonic nature of the free spin. Furthermore, the applicability of the lowest order expansion itself for the description of the spin-dependent IETS experiments seems to be currently under debate and renormalized perturbation theories might be necessary for a proper description of observed phenomena. Addressing
these problems in the noise context constitutes an interesting future research direction.

Finally, the calculation of arbitrary cumulants based on the generalized Keldysh GF technique can be implemented numerically~\cite{Fransson,Urban} to address the cases of structured tunneling density of states and/or stronger $e$-ph
coupling, which go beyond our analytical treatment. On the other
hand, for the case of weak coupling addressed in this work, such numerical methods 
will face convergence/efficiency problems due to very sharp
phonon lineshapes and unnecessary self-consistency loops. In this respect, when complemented
by {\em ab-initio} calculations for the transport coefficients $\Tr{\Tlr},\Tr{\Tlr_{H}^{\rm (qel)}},\dots$,
our approach is designed to be a very efficient alternative to the full numerics in the  
limit of weak coupling and slowly varying electronic density of states. It uses the realistic static
calculations of the electronic Green's functions, phonon modes, and
their coupling as input parameters and yields reliable results for
the dynamical effects in the electronic noise.

\begin{acknowledgments}
We  thank D.~Bagrets, A.-P.~Jauho, D.~F.~Urban, and J.~M.~van Ruitenbeek for useful discussions, Yu.~V.~Nazarov for providing us with Ref.~\onlinecite{Jouravlev},  and A.~Braggio for invaluable help
with {\tt Mathematica}. We acknowledge the financial support by
DFG via SFB 767 (F.~H. and W.~B.), by the Czech Science Foundation
via the grant 202/07/J051 and the Ministry of Education of the
Czech Republic via the research plan MSM 0021620834 (T.~N.).
\end{acknowledgments}

\appendix

\section{Mean-field contribution to noise}\label{App:MeanField}
Using the invariance of the trace under cyclic permutations,
Eq.~\eqref{eq:Smf} can be recast in the following form
\begin{equation}\label{eq:Smf:app}
\begin{split}
\frac{S^{\rm (mf)}}{e^{2}/h}&=\int d \e \,{\rm Tr} \big\{ i\Ga_{L}\big(f_{L}\G^{>}-(1-f_{L})\G^{<}\big) \\
    &+\Ga_{L}\G^{>}\Ga_{L}\G^{<}+\Ga_{L}(\G^{r}-\G^{a})\Ga_{L}\big(f_{L}\G^{>}\\
    &-(1-f_{L})\G^{<}\big)-f_{L}(1-f_{L})\big( \G^{a}\Ga_{L}\G^{a}\Ga_{L}\\
    &+\G^{r}\Ga_{L}\G^{r}\Ga_{L}\big)\big\},
\end{split}
\end{equation}
with $\G^{\lessgtr}\equiv \G^{\mp\pm}_{\bl=0}$, $\G^{r}=\G^{--}_{\bl=0}-\G^{-+}_{\bl=0}$ and $\G^{a}=[\G^{r}]^{\dag}$. Such an expression
corresponds exactly~\footnote{Apart from an extra factor of 2 in
Eqs. (30), (31) of Ref.~\onlinecite{Souza} stemming from their
different definition of the zero-frequency noise.} to Eq. (30) of
Ref.~\onlinecite{Souza}. We stress that their result was obtained
by truncating the $S$-matrix expansion by breaking two-particle
Green's functions into products of one-particle Green's functions,
see Refs.~\onlinecite{Souza} or \onlinecite{Jauho:book}--Sec.~13.8  for
further details. This procedure holds in a mean-field theory, but
it misses the contributions given by the vertex correction.  For
this reason we identified Eq.~\eqref{eq:Smf} with the {\em
mean-field contribution} to noise.

It can be furthermore shown that Eq.~\eqref{eq:Smf:app} is 
equivalent to Eq.~(10) of Ref.~\onlinecite{Zhu}
and 
to  the zero-frequency limit of Eq.~(9) of Ref.~\onlinecite{Galperin:PRB06}, 
which therefore represent again {\em solely} the mean-field contribution to the noise.
We stress however that, as we will discuss in Appendix ~\ref{App:FDT}, approximating the noise with the
mean-field contribution generally leads to violation of the 
fluctuation-dissipation theorem. 


\section{Electronic density}\label{App:ne}
The electronic density in the system is given by
\begin{equation} \label{eq:ne}
\n_{e}=-i\int\frac{d\e}{2\pi}\g^{- +}_{\bl=0}(\e),
\end{equation}
where $\g^{- +}_{\bl=0}=\g^{r}\Si_{T}^{-+}\g^{a}|_{\bl=0}$ is the
lesser Green's function  without the electron-phonon coupling. It
should be noticed that the integrand of Eq.~\eqref{eq:ne} does
{\em not} have a finite support and therefore in this case
integration over energy cannot be carried out in the eWBL
approximation. Instead, the energy dependence of $\g^{r(a)}$ has
to be taken into account while calculating the integral, and only
subsequently one is allowed to consider the limits  $\Gamma \gg
eV,  \hbar \om_{0}, k_{B}T$  or $|\e_{\rm res}-E_{F}|\gg \Gamma, eV, \hbar
\om_{0}, k_{B}T$ corresponding to the eWBL.

As an example we consider here the case of a system with a single
electronic level symmetrically coupled to unstructured leads with constant $\Gamma_{L}=\Gamma_{R}\equiv\Gamma$. In this case
\begin{equation}
n_{e}=-i\int\frac{d\e}{2\pi}\frac{\Gamma[f_{L}(\e)+f_{R}(\e)]}{(\e-\e_{0})^{2}+\Gamma^{2}}.
\end{equation}
Assuming zero temperature and symmetric voltage drop at the
barriers $\mu_{L}=-\mu_{R}=eV/2$ one gets
\begin{equation*}
\begin{split}
\n_{e}&=\frac{1}{2}+\frac{1}{2\pi}\left[\arctan\left(\! \frac{eV-2\e_{0}}{2\Gamma} \!\right)-\arctan\left(\!\frac{eV+2\e_{0}}{2 \Gamma}\! \right)  \right]\\
    &=\frac{1}{2}-\frac{1}{\pi}\arctan\left(\frac{\e_{0}}{\Gamma}\right)+\frac{\Gamma \e_{0} }{4\pi (\Gamma^{2}+\e_{0}^{2})^{2}}(eV)^{2}+\mathcal{O}((eV)^{3}),
\end{split}
\end{equation*}
which shows that in the eWBL $\n_{e}$ depends very weakly on the
applied bias voltage.


\section{Explicit form of the Fock self-energy at zero counting field}\label{App:selfenergy}
At $\bl=0$,  the Keldysh components of $\check{\Si}_{F}$ satisfy
the identity
$(\Si_{F}^{--}+\Si_{F}^{++})_{\bl=0}=(\Si_{F}^{-+}+\Si_{F}^{+-})_{\bl=0}$.
In this case, it is meaningful to introduce the retarded and
advanced self-energies by  $\Si_{F}^{r}\equiv
(\Si_{F}^{--}-\Si_{F}^{-+})_{\bl=0}$  and $\Si_{F}^{a}=\big[
\Si_{F}^{r} \big]^{\dag}$. For definiteness, we also introduce the
notation $\Si^{\lessgtr}_{F}\equiv \Si^{\mp\pm}_{\bl=0}$ for the
lesser and greater components at zero counting field. The latter
can be easily calculated from Eq.~\eqref{eq:SigmaF} giving
\begin{align*}
\Si^{<}_{F}(\e)=&\phantom{-}i \!\sum_{\alpha=L,R} \! \M \big[\np  \A_{\alpha}(\e-\hbar\om_0) f_{\alpha}(\e-\hbar\om_0)\\
    &+(\np+1)\A_{\alpha}(\e+\hbar\om_0) f_{\alpha}(\e+\hbar\om_0) \big]\M \\
\Si^{>}_{F}(\e)=&-i\!\sum_{\alpha=L,R} \!\M \big[\np  \A_{\alpha}(\e+\hbar\om_0)\big(1- f_{\alpha}(\e+\hbar\om_0)\big)\\
    &+ (\np+1)\A_{\alpha}(\e-\hbar\om_0)\big(1- f_{\alpha}(\e-\hbar\om_0)\big)\big]\M
\end{align*}
The retarded self-energy can in turn be written in terms of the lesser
and greater components using the identity
$\Si_{F}^{r}-\Si_{F}^{a}=\Si_{F}^{>}-\Si_{F}^{<}$  and
Kramers-Kronig relation ${\rm
Re}\Si_{F}^{r}(\e)=\mathcal{H}_{\e'}\big\{{\rm Im}\Si_{F}^{r}(\e')
\big\}(\e)$. This leads to
\begin{align*}
{\rm Im}\Si^r(\e)=&-\!\frac{1}{2}\M \big\{\!(\np\!+\!1)\A(\e\!-\! \hbar\om_0)\!+\!\np \A(\!\e+\! \hbar\om_0)\!\big\} \M\\
      &-\frac{1}{2}\sum_{\alpha=L,R}\M\big\{ \A_{\alpha}(\e+\hbar\om_0)  f_{\alpha}(\e+\hbar\om_0)\\
      &-\A_{\alpha}(\e-\hbar\om_0)  f_{\alpha}(\e-\hbar\om_0)\big\}\M,
\end{align*}
\begin{align*}
{\rm Re}\Si^r(\e)=&\M\big[(\np+1)\g^{r}_{R}(\e-\hbar\om_0)+\np \g^{r}_{R}(\e+\hbar\om_0)\big]\M\\
      &-\frac{1}{2}\sum_{\alpha=L,R}\M\big[ \Ht\{\A_{\alpha}(\e') f_{\alpha}(\e')\}(\e+\hbar\om_0)\\
      &-\Ht\{\A_{\alpha}(\e')  f_{\alpha}(\e')\}(\e-\hbar\om_0)\big]\M,
\end{align*}
where we have used the identity $\Ht\{\A(\e')\}(\e)=-2 \Ht\{
\g^r_{I}(\e') \}(\e)=-2 \g^r_{R}(\e)$, with $\g^{r}_{R(I)}$ the
real (imaginary) part of $\g^{r}$. 
We point out that the energy dependence of $\g^{r(a)}$ cannot be disregarded 
while using Kramers-Kronig relations, as the Hilbert transform $\Ht$ generally 
involves integrals over infinitely extended range. However, in the limits  $\Gamma \gg
eV,  \hbar \om_{0},k_{B}T$ or $|\e_{\rm res}-E_{F}|\gg \Gamma, eV, \hbar
\om_{0}, k_{B}T$ corresponding to the eWBL approximation
the previous expressions take a simpler form
\begin{equation}\label{eq:ImSiF}
\begin{split}
{\rm Im}\Si^r_{F}(\e)=&-\frac{1}{2}\sum_{\alpha=L,R}\M \A_{\alpha}\M\big[(2\np+1) \\
    &+f_{\alpha}(\e+\hbar\om_0)- f_{\alpha}(\e-\hbar\om_0) \big],
\end{split}
\end{equation}
\begin{equation}\label{eq:ReSiF}
\begin{split}
{\rm Re}\Si^r_{F}(\e)&=(2\np+1)\M\g^{r}_{R}\M -\frac{1}{2}\sum_{\alpha=L,R}\M\A_{\alpha}\M \\
 &\times \Ht\{f_{\alpha}(\e'+\hbar\om_0)-  f_{\alpha}(\e'-\hbar\om_0)\}(\e).
\end{split}
\end{equation}
Note that the Hilbert transform now involves only a function with finite support.
Inserting these expressions into Eq.~\eqref{eq:I_FH}, it is easy to show
that the first term of Eq.~\eqref{eq:ReSiF} is the origin of the discrepancy between our
result for $I_{F}$, Eq.~\eqref{eq:I_{F}}, and the expression
derived by Paulsson and
coworkers.~\cite{Paulsson:RapCom,Frederiksen:PRB07}  This discrepancy 
stems from the subtleties in the use of the eWBL mentioned above, 
and was already pointed out by Viljas {\em et al}.,~\cite{Viljas} whose result agrees with ours.

\section{Fluctuation-dissipation theorem}\label{App:FDT}
The fluctuation-dissipation theorem relates the noise at zero
voltage to the linear conductance of the system  $G$
\begin{equation}\nonumber
S(V=0)=\frac{2}{\beta}G.
\end{equation}

In the case of the contributions due to the Hartree term, it follows form Eqs.~\eqref{eq:SHvc},~\eqref{eq:SHmf} that   
$S_{H}^{\rm (vc)}\to 0$ at zero voltage, while $S_{H}^{\rm (mf)}$ fulfills the fluctuation-dissipation theorem 
$$S_{H}(V=0)=S_{H}^{\rm (mf)}(V=0)=\frac{2}{\beta}G_{H},$$
with $G_{H}=e^{2}/h\Tr{\Tlr_{H}^{\rm (qel)}}$.   

The situation is, however, different for the Fock term. In fact, 
in the limit $eV\to 0$, both the mean-field contribution $S_{F}^{\rm (mf)}$ and the vertex
corrections  $S_{F}^{\rm (vc)}$ are different from zero and reduce to
\begin{equation}\label{eq:SFmf0}
\begin{split} \nonumber
\frac{S_{F}^{\rm (mf)}(V=0)}{e^{2}/h}&=\Tr{\Tlr^{\rm(qel)}_{F}}\frac{2}{\beta}\frac{U(\hbar\om_{0})}{\hbar\om_{0}}+\frac{U(\hbar\om_{0})^{2}\!-(\hbar\om_{0})^{2}}{\hbar\om_{0}}\\
&\times \Tr{\Tlr_{F}^{\rm (inel)} +\Ga_{L}\g^{r}\M\A_{L}\M\g^{a}},
\end{split}
\end{equation}
\begin{equation}\label{eq:SFvc0}
\begin{split}\nonumber
\frac{S_{F}^{\rm
(vc)}(V=0)}{e^{2}/h}&=-\Tr{\Ga_{L}\g^{r}\M\A_{L}\M\g^{a}}\frac{U(\hbar\om_{0})^{2}\!-(\hbar\om_{0})^{2}}{\hbar\om_{0}}.
\end{split}
\end{equation}
On the other hand, the correction to the linear conductance due to
$\check{\Si}_{F}$ is given by
\begin{equation}\nonumber
\frac{G_{F}}{e^{2}/h}\!=\!\Tr{\Tlr^{\rm(qel)}_{F}}\frac{U(\hbar\om_{0})}{\hbar\om_{0}}+\!\Tr{\Tlr^{\rm(inel)}_{F}}\frac{\beta}{2}\frac{U(\hbar\om_{0})^{2}\!-(\hbar\om_{0})^{2}}{\hbar\om_{0}}.
\end{equation}
Comparing the previous expressions one can see that 
\begin{equation} \nonumber
S_{F}^{\rm (mf)}(V=0)+S_{F}^{\rm (vc)}(V=0)=\frac{2}{\beta}G_{F}.
\end{equation}
but the mean field contribution alone does not satisfy the 
fluctuation dissipation theorem $S_{F}^{\rm (mf)}(V=0)\neq2 G_{F}/\beta$.
This clearly shows that in general, even in the limit of weak $e$-ph
coupling, vertex corrections {\em must}  be included into the
noise calculation in order to obtain consistent results.


\section{Anti-symmetric contribution to $S_{F}$}
\label{App:asym}
\begin{figure}
\begin{center}
{\resizebox{0.85\columnwidth}{!}{\includegraphics{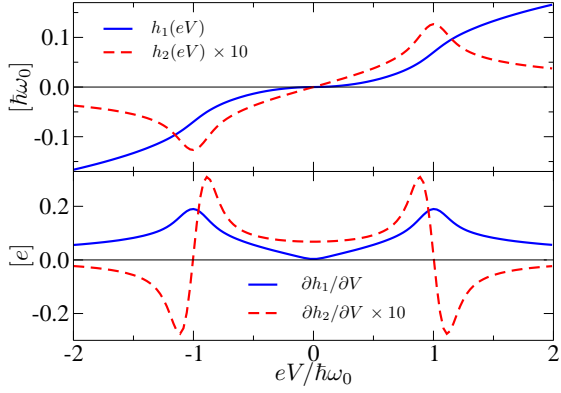}}}
\end{center}
\caption{Upper panel: Plots of the dependencies of the functions $h_{1}(eV)$ and $h_{2}(eV)$ on the applied bias voltage (Eqs.~(\ref{eq:h1}) and (\ref{eq:h2}), respectively). 
Lower panel: Same as above, but for the derivatives $\partial h_{1}/\partial V$ and $\partial h_{2}/\partial V$. In both panels  $k_{B}T=\hbar\om_{0}/30$.
}\label{fig:plot:h1h2}   
\end{figure}
In this appendix we give the complete expression for the terms of $S_{F}^{\rm (mf/vc)}$
which are anti-symmetric with respect to the bias voltage. At finite temperature 
they are given by
\begin{subequations}
\begin{align}
\frac{S_{\rm asym}^{(\rm mf)}}{e^{2}/h}=\Tr{\K_{1}^{\rm (mf)}}h_{1}(eV)+\Tr{\K_{2}^{(\rm mf)}}h_{2}(eV),\\
\frac{S_{\rm asym}^{(\rm vc)}}{e^{2}/h}=\Tr{\K_{1}^{\rm (vc)}}h_{1}(eV)+\Tr{\K_{2}^{(\rm vc)}}h_{2}(eV),
\end{align}
\end{subequations}
where $\K_{1}^{\rm (mf/vc)}$ are those of Eqs.~\eqref{eq:K1} and 
\begin{subequations}
\begin{align}
\K_{2}^{(\rm mf)}&=\Ga_{L}\A_{R}\Ga_{L}[\A_{R}\M\,(\A_{L}-\A_{R})\, \g^{a}+h.c.],\\
\K_{2}^{(\rm vc)}&= i\M(\A_{R}\Ga_{L}\g^{r}\!+h.c.)\M(\A \Ga_{L}\g_{R}^{r}\!-h.c.)-\K_{1}^{(\rm vc)}/2. 
\end{align}
\end{subequations}
The line shape of  $S_{\rm asym}^{(\rm mf/vc)}$ is defined by the functions 
\begin{equation}\label{eq:h1}
h_{1}(eV)=h(eV)\coth(\beta \, eV/2),
\end{equation}
with $h(eV)$ given in Eq.~\eqref{eq:h} and
\begin{equation}\label{eq:h2}
\begin{split}
h_{2}(eV)=&\int d\e\, \Big[ n_{F}(\e+eV)[1-n_{F}(\e+eV)]\\
	&\times\Ht\big\{n_{F}(\e'-\hbar\om_{0})-n_{F}(\e'+\hbar\om_{0}) \big\}(\e)\Big].
\end{split}
\end{equation}
We notice that  $h_{2}(eV)=(e\beta)^{-1}\partial_{V} h(eV)$, i.e.\ $h_{2}(eV)$ is directly proportional to the derivative of $h(eV)$
but it exhibits no divergencies and actually tends to zero in the limit $T\to 0$ due to the suppression factor $1/\beta$ with respect 
to $h(eV)$. Furthermore, we observe once again that  
$S_{\rm asym}^{(\rm mf/vc)}$ give a negligible contribution to the inelastic noise signal $\Delta S'$ at low temperatures, since $\partial_{V} h_{1}$ is almost symmetric around the phonon emission threshold, i.e. $\partial_{V} h_{1}\big|_{|eV|\gtrsim\hbar\om_{0}}\approx \partial_{V} h_{1}\big|_{|eV|\lesssim\hbar\om_{0}}$, and $\partial_{V} h_{2}$ is suppressed by low temperature, see Fig.~\ref{fig:plot:h1h2}.



\begin{thebibliography}{80}%
\makeatletter
\providecommand \@ifxundefined [1]{%
 \@ifx{#1\undefined}
}%
\providecommand \@ifnum [1]{%
 \ifnum #1\expandafter \@firstoftwo
 \else \expandafter \@secondoftwo
 \fi
}%
\providecommand \@ifx [1]{%
 \ifx #1\expandafter \@firstoftwo
 \else \expandafter \@secondoftwo
 \fi
}%
\providecommand \natexlab [1]{#1}%
\providecommand \enquote  [1]{``#1''}%
\providecommand \bibnamefont  [1]{#1}%
\providecommand \bibfnamefont [1]{#1}%
\providecommand \citenamefont [1]{#1}%
\providecommand \href@noop [0]{\@secondoftwo}%
\providecommand \href [0]{\begingroup \@sanitize@url \@href}%
\providecommand \@href[1]{\@@startlink{#1}\@@href}%
\providecommand \@@href[1]{\endgroup#1\@@endlink}%
\providecommand \@sanitize@url [0]{\catcode `\\12\catcode `\$12\catcode
  `\&12\catcode `\#12\catcode `\^12\catcode `\_12\catcode `\%12\relax}%
\providecommand \@@startlink[1]{}%
\providecommand \@@endlink[0]{}%
\providecommand \url  [0]{\begingroup\@sanitize@url \@url }%
\providecommand \@url [1]{\endgroup\@href {#1}{\urlprefix }}%
\providecommand \urlprefix  [0]{URL }%
\providecommand \Eprint [0]{\href }%
\@ifxundefined \urlstyle {%
  \providecommand \doi  [0]{\begingroup \@sanitize@url \@doi}%
  \providecommand \@doi [1]{\endgroup \@@startlink {\doibase
  #1}doi:\discretionary {}{}{}#1\@@endlink }%
}{%
  \providecommand \doi  [0]{doi:\discretionary{}{}{}\begingroup
  \urlstyle{rm}\Url }%
}%
\providecommand \doibase [0]{http://dx.doi.org/}%
\providecommand \Doi [0]{\begingroup \@sanitize@url \@Doi }%
\providecommand \@Doi  [1]{\endgroup\@@startlink{\doibase#1}\@@Doi}%
\providecommand \@@Doi [1]{#1\@@endlink}%
\providecommand \selectlanguage [0]{\@gobble}%
\providecommand \bibinfo  [0]{\@secondoftwo}%
\providecommand \bibfield  [0]{\@secondoftwo}%
\providecommand \translation [1]{[#1]}%
\providecommand \BibitemOpen [0]{}%
\providecommand \bibitemStop [0]{}%
\providecommand \bibitemNoStop [0]{.\EOS\space}%
\providecommand \EOS [0]{\spacefactor3000\relax}%
\providecommand \BibitemShut  [1]{\csname bibitem#1\endcsname}%
\bibitem [{\citenamefont {Cuniberti}\ \emph {et~al.}(2005)\citenamefont
  {Cuniberti}, \citenamefont {Fagas},\ and\ \citenamefont
  {Richter}}]{Cuniberti}%
  \BibitemOpen
  \bibinfo {editor} {\bibfnamefont {G.}~\bibnamefont {Cuniberti}}, \bibinfo
  {editor} {\bibfnamefont {G.}~\bibnamefont {Fagas}}, \ and\ \bibinfo {editor}
  {\bibfnamefont {K.}~\bibnamefont {Richter}},\ eds.,\ \href@noop {} {\emph
  {\bibinfo {title} {Introducing Molecular Electronics}}}\ (\bibinfo
  {publisher} {Springer},\ \bibinfo {address} {Berlin},\ \bibinfo {year}
  {2005})\BibitemShut {NoStop}%
\bibitem [{\citenamefont {Agra\"\i{}t}\ \emph {et~al.}(2003)\citenamefont
  {Agra\"\i{}t}, \citenamefont {Yeyati},\ and\ \citenamefont {van
  Ruitenbeek}}]{Agrait:PhysRep}%
  \BibitemOpen
  \bibfield  {author} {\bibinfo {author} {\bibfnamefont {N.}~\bibnamefont
  {Agra\"\i{}t}}, \bibinfo {author} {\bibfnamefont {A.~L.}\ \bibnamefont
  {Yeyati}}, \ and\ \bibinfo {author} {\bibfnamefont {J.~M.}\ \bibnamefont {van
  Ruitenbeek}},\ }\Doi {DOI: 10.1016/S0370-1573(02)00633-6} {\bibfield
  {journal} {\bibinfo  {journal} {Phys. Rep.},\ }\textbf {\bibinfo {volume}
  {377}},\ \bibinfo {pages} {81 } (\bibinfo {year} {2003})},\ ISSN \bibinfo
  {issn} {0370-1573}\BibitemShut {NoStop}%
\bibitem [{\citenamefont {Naidyuk}\ and\ \citenamefont
  {Yanson}(2005)}]{Naidyuk}%
  \BibitemOpen
  \bibfield  {author} {\bibinfo {author} {\bibfnamefont {Y.~G.}\ \bibnamefont
  {Naidyuk}}\ and\ \bibinfo {author} {\bibfnamefont {I.~K.}\ \bibnamefont
  {Yanson}},\ }\href@noop {} {\emph {\bibinfo {title} {Point-Contact
  Spectroscopy}}}\ (\bibinfo  {publisher} {Springer},\ \bibinfo {address}
  {Berlin},\ \bibinfo {year} {2005})\BibitemShut {NoStop}%
\bibitem [{\citenamefont {Jaklevic}\ and\ \citenamefont
  {Lambe}(1966)}]{Jaklevic}%
  \BibitemOpen
  \bibfield  {author} {\bibinfo {author} {\bibfnamefont {R.~C.}\ \bibnamefont
  {Jaklevic}}\ and\ \bibinfo {author} {\bibfnamefont {J.}~\bibnamefont
  {Lambe}},\ }\Doi {10.1103/PhysRevLett.17.1139} {\bibfield  {journal}
  {\bibinfo  {journal} {Phys. Rev. Lett.},\ }\textbf {\bibinfo {volume} {17}},\
  \bibinfo {pages} {1139} (\bibinfo {year} {1966})}\BibitemShut {NoStop}%
\bibitem [{\citenamefont {Stipe}\ \emph {et~al.}(1998)\citenamefont {Stipe},
  \citenamefont {Rezaei},\ and\ \citenamefont {Ho}}]{Stipe}%
  \BibitemOpen
  \bibfield  {author} {\bibinfo {author} {\bibfnamefont {B.~C.}\ \bibnamefont
  {Stipe}}, \bibinfo {author} {\bibfnamefont {M.~A.}\ \bibnamefont {Rezaei}}, \
  and\ \bibinfo {author} {\bibfnamefont {W.}~\bibnamefont {Ho}},\ }\Doi
  {10.1126/science.280.5370.1732} {\bibfield  {journal} {\bibinfo  {journal}
  {Science},\ }\textbf {\bibinfo {volume} {280}},\ \bibinfo {pages} {1732}
  (\bibinfo {year} {1998})}\BibitemShut {NoStop}%
\bibitem [{\citenamefont {Agra\"\i{}t}\ \emph {et~al.}(2002)\citenamefont
  {Agra\"\i{}t}, \citenamefont {Untiedt}, \citenamefont {Rubio-Bollinger},\
  and\ \citenamefont {Vieira}}]{Agrait:PRL02}%
  \BibitemOpen
  \bibfield  {author} {\bibinfo {author} {\bibfnamefont {N.}~\bibnamefont
  {Agra\"\i{}t}}, \bibinfo {author} {\bibfnamefont {C.}~\bibnamefont
  {Untiedt}}, \bibinfo {author} {\bibfnamefont {G.}~\bibnamefont
  {Rubio-Bollinger}}, \ and\ \bibinfo {author} {\bibfnamefont {S.}~\bibnamefont
  {Vieira}},\ }\Doi {10.1103/PhysRevLett.88.216803} {\bibfield  {journal}
  {\bibinfo  {journal} {Phys. Rev. Lett.},\ }\textbf {\bibinfo {volume} {88}},\
  \bibinfo {pages} {216803} (\bibinfo {year} {2002})}\BibitemShut {NoStop}%
\bibitem [{\citenamefont {Smit}\ \emph {et~al.}(2002)\citenamefont {Smit},
  \citenamefont {Noat}, \citenamefont {Untiedt}, \citenamefont {Lang},
  \citenamefont {van Hemert},\ and\ \citenamefont {van Ruitenbeek}}]{Smit}%
  \BibitemOpen
  \bibfield  {author} {\bibinfo {author} {\bibfnamefont {R.~H.~M.}\
  \bibnamefont {Smit}}, \bibinfo {author} {\bibfnamefont {Y.}~\bibnamefont
  {Noat}}, \bibinfo {author} {\bibfnamefont {C.}~\bibnamefont {Untiedt}},
  \bibinfo {author} {\bibfnamefont {N.~D.}\ \bibnamefont {Lang}}, \bibinfo
  {author} {\bibfnamefont {M.~C.}\ \bibnamefont {van Hemert}}, \ and\ \bibinfo
  {author} {\bibfnamefont {J.~M.}\ \bibnamefont {van Ruitenbeek}},\ }\Doi
  {10.1038/nature01103} {\bibfield  {journal} {\bibinfo  {journal} {Nature},\
  }\textbf {\bibinfo {volume} {419}},\ \bibinfo {pages} {906} (\bibinfo {year}
  {2002})}\BibitemShut {NoStop}%
\bibitem [{\citenamefont {Djukic}\ \emph {et~al.}(2005)\citenamefont {Djukic},
  \citenamefont {Thygesen}, \citenamefont {Untiedt}, \citenamefont {Smit},
  \citenamefont {Jacobsen},\ and\ \citenamefont {van Ruitenbeek}}]{Djukic_PRB}%
  \BibitemOpen
  \bibfield  {author} {\bibinfo {author} {\bibfnamefont {D.}~\bibnamefont
  {Djukic}}, \bibinfo {author} {\bibfnamefont {K.~S.}\ \bibnamefont
  {Thygesen}}, \bibinfo {author} {\bibfnamefont {C.}~\bibnamefont {Untiedt}},
  \bibinfo {author} {\bibfnamefont {R.~H.~M.}\ \bibnamefont {Smit}}, \bibinfo
  {author} {\bibfnamefont {K.~W.}\ \bibnamefont {Jacobsen}}, \ and\ \bibinfo
  {author} {\bibfnamefont {J.~M.}\ \bibnamefont {van Ruitenbeek}},\ }\Doi
  {10.1103/PhysRevB.71.161402} {\bibfield  {journal} {\bibinfo  {journal}
  {Phys. Rev. B},\ }\textbf {\bibinfo {volume} {71}},\ \bibinfo {pages}
  {161402} (\bibinfo {year} {2005})}\BibitemShut {NoStop}%
\bibitem [{\citenamefont {Tal}\ \emph {et~al.}(2008)\citenamefont {Tal},
  \citenamefont {Krieger}, \citenamefont {Leerink},\ and\ \citenamefont {van
  Ruitenbeek}}]{Tal}%
  \BibitemOpen
  \bibfield  {author} {\bibinfo {author} {\bibfnamefont {O.}~\bibnamefont
  {Tal}}, \bibinfo {author} {\bibfnamefont {M.}~\bibnamefont {Krieger}},
  \bibinfo {author} {\bibfnamefont {B.}~\bibnamefont {Leerink}}, \ and\
  \bibinfo {author} {\bibfnamefont {J.~M.}\ \bibnamefont {van Ruitenbeek}},\
  }\Doi {10.1103/PhysRevLett.100.196804} {\bibfield  {journal} {\bibinfo
  {journal} {Phys. Rev. Lett.},\ }\textbf {\bibinfo {volume} {100}},\ \bibinfo
  {pages} {196804} (\bibinfo {year} {2008})}\BibitemShut {NoStop}%
\bibitem [{\citenamefont {Frederiksen}\ \emph {et~al.}(2008)\citenamefont
  {Frederiksen}, \citenamefont {Franke}, \citenamefont {Arnau}, \citenamefont
  {Schulze}, \citenamefont {Pascual},\ and\ \citenamefont {Lorente}}]{Franke}%
  \BibitemOpen
  \bibfield  {author} {\bibinfo {author} {\bibfnamefont {T.}~\bibnamefont
  {Frederiksen}}, \bibinfo {author} {\bibfnamefont {K.~J.}\ \bibnamefont
  {Franke}}, \bibinfo {author} {\bibfnamefont {A.}~\bibnamefont {Arnau}},
  \bibinfo {author} {\bibfnamefont {G.}~\bibnamefont {Schulze}}, \bibinfo
  {author} {\bibfnamefont {J.~I.}\ \bibnamefont {Pascual}}, \ and\ \bibinfo
  {author} {\bibfnamefont {N.}~\bibnamefont {Lorente}},\ }\Doi
  {10.1103/PhysRevB.78.233401} {\bibfield  {journal} {\bibinfo  {journal}
  {Phys. Rev. B},\ }\textbf {\bibinfo {volume} {78}},\ \bibinfo {pages}
  {233401} (\bibinfo {year} {2008})}\BibitemShut {NoStop}%
\bibitem [{\citenamefont {Rahimi}\ and\ \citenamefont
  {Hegg}(2009)}]{Rahimi:2009}%
  \BibitemOpen
  \bibfield  {author} {\bibinfo {author} {\bibfnamefont {M.}~\bibnamefont
  {Rahimi}}\ and\ \bibinfo {author} {\bibfnamefont {M.}~\bibnamefont {Hegg}},\
  }\Doi {10.1103/PhysRevB.79.081404} {\bibfield  {journal} {\bibinfo  {journal}
  {Phys. Rev. B},\ }\textbf {\bibinfo {volume} {79}},\ \bibinfo {pages}
  {081404} (\bibinfo {year} {2009})}\BibitemShut {NoStop}%
\bibitem [{\citenamefont {Arroyo}\ \emph {et~al.}(2010)\citenamefont {Arroyo},
  \citenamefont {Frederiksen}, \citenamefont {Rubio-Bollinger}, \citenamefont
  {V\'elez}, \citenamefont {Arnau}, \citenamefont {S\'anchez-Portal},\ and\
  \citenamefont {Agra\"\i{}t}}]{Arroyo:2010}%
  \BibitemOpen
  \bibfield  {author} {\bibinfo {author} {\bibfnamefont {C.~R.}\ \bibnamefont
  {Arroyo}}, \bibinfo {author} {\bibfnamefont {T.}~\bibnamefont {Frederiksen}},
  \bibinfo {author} {\bibfnamefont {G.}~\bibnamefont {Rubio-Bollinger}},
  \bibinfo {author} {\bibfnamefont {M.}~\bibnamefont {V\'elez}}, \bibinfo
  {author} {\bibfnamefont {A.}~\bibnamefont {Arnau}}, \bibinfo {author}
  {\bibfnamefont {D.}~\bibnamefont {S\'anchez-Portal}}, \ and\ \bibinfo
  {author} {\bibfnamefont {N.}~\bibnamefont {Agra\"\i{}t}},\ }\Doi
  {10.1103/PhysRevB.81.075405} {\bibfield  {journal} {\bibinfo  {journal}
  {Phys. Rev. B},\ }\textbf {\bibinfo {volume} {81}},\ \bibinfo {pages}
  {075405} (\bibinfo {year} {2010})}\BibitemShut {NoStop}%
\bibitem [{\citenamefont {Heinrich}\ \emph {et~al.}(2004)\citenamefont
  {Heinrich}, \citenamefont {Gupta}, \citenamefont {Lutz},\ and\ \citenamefont
  {Eigler}}]{Eigler}%
  \BibitemOpen
  \bibfield  {author} {\bibinfo {author} {\bibfnamefont {A.~J.}\ \bibnamefont
  {Heinrich}}, \bibinfo {author} {\bibfnamefont {J.~A.}\ \bibnamefont {Gupta}},
  \bibinfo {author} {\bibfnamefont {C.~P.}\ \bibnamefont {Lutz}}, \ and\
  \bibinfo {author} {\bibfnamefont {D.~M.}\ \bibnamefont {Eigler}},\ }\Doi
  {10.1126/science.1101077} {\bibfield  {journal} {\bibinfo  {journal}
  {Science},\ }\textbf {\bibinfo {volume} {306}},\ \bibinfo {pages} {466}
  (\bibinfo {year} {2004})}\BibitemShut {NoStop}%
\bibitem [{\citenamefont {Hirjibehedin}\ \emph {et~al.}(2006)\citenamefont
  {Hirjibehedin}, \citenamefont {Lutz},\ and\ \citenamefont
  {Heinrich}}]{Heinrich}%
  \BibitemOpen
  \bibfield  {author} {\bibinfo {author} {\bibfnamefont {C.~F.}\ \bibnamefont
  {Hirjibehedin}}, \bibinfo {author} {\bibfnamefont {C.~P.}\ \bibnamefont
  {Lutz}}, \ and\ \bibinfo {author} {\bibfnamefont {A.~J.}\ \bibnamefont
  {Heinrich}},\ }\Doi {10.1126/science.1125398} {\bibfield  {journal} {\bibinfo
   {journal} {Science},\ }\textbf {\bibinfo {volume} {312}},\ \bibinfo {pages}
  {1021} (\bibinfo {year} {2006})}\BibitemShut {NoStop}%
\bibitem [{\citenamefont {Fu}\ \emph {et~al.}(2009)\citenamefont {Fu},
  \citenamefont {Zhang}, \citenamefont {Ji}, \citenamefont {Chen},
  \citenamefont {Ma}, \citenamefont {Jia},\ and\ \citenamefont {Xue}}]{Cinane}%
  \BibitemOpen
  \bibfield  {author} {\bibinfo {author} {\bibfnamefont {Y.-S.}\ \bibnamefont
  {Fu}}, \bibinfo {author} {\bibfnamefont {T.}~\bibnamefont {Zhang}}, \bibinfo
  {author} {\bibfnamefont {S.-H.}\ \bibnamefont {Ji}}, \bibinfo {author}
  {\bibfnamefont {X.}~\bibnamefont {Chen}}, \bibinfo {author} {\bibfnamefont
  {X.-C.}\ \bibnamefont {Ma}}, \bibinfo {author} {\bibfnamefont {J.-F.}\
  \bibnamefont {Jia}}, \ and\ \bibinfo {author} {\bibfnamefont {Q.-K.}\
  \bibnamefont {Xue}},\ }\Doi {10.1103/PhysRevLett.103.257202} {\bibfield
  {journal} {\bibinfo  {journal} {Phys. Rev. Lett.},\ }\textbf {\bibinfo
  {volume} {103}},\ \bibinfo {pages} {257202} (\bibinfo {year}
  {2009})}\BibitemShut {NoStop}%
\bibitem [{\citenamefont {Lorente}\ and\ \citenamefont
  {Persson}(2000)}]{LorentePersson}%
  \BibitemOpen
  \bibfield  {author} {\bibinfo {author} {\bibfnamefont {N.}~\bibnamefont
  {Lorente}}\ and\ \bibinfo {author} {\bibfnamefont {M.}~\bibnamefont
  {Persson}},\ }\Doi {10.1103/PhysRevLett.85.2997} {\bibfield  {journal}
  {\bibinfo  {journal} {Phys. Rev. Lett.},\ }\textbf {\bibinfo {volume} {85}},\
  \bibinfo {pages} {2997} (\bibinfo {year} {2000})}\BibitemShut {NoStop}%
\bibitem [{\citenamefont {Frederiksen}\ \emph {et~al.}(2004)\citenamefont
  {Frederiksen}, \citenamefont {Brandbyge}, \citenamefont {Lorente},\ and\
  \citenamefont {Jauho}}]{Frederiksen:PRL04}%
  \BibitemOpen
  \bibfield  {author} {\bibinfo {author} {\bibfnamefont {T.}~\bibnamefont
  {Frederiksen}}, \bibinfo {author} {\bibfnamefont {M.}~\bibnamefont
  {Brandbyge}}, \bibinfo {author} {\bibfnamefont {N.}~\bibnamefont {Lorente}},
  \ and\ \bibinfo {author} {\bibfnamefont {A.-P.}\ \bibnamefont {Jauho}},\
  }\Doi {10.1103/PhysRevLett.93.256601} {\bibfield  {journal} {\bibinfo
  {journal} {Phys. Rev. Lett.},\ }\textbf {\bibinfo {volume} {93}},\ \bibinfo
  {pages} {256601} (\bibinfo {year} {2004})}\BibitemShut {NoStop}%
\bibitem [{\citenamefont {Paulsson}\ \emph {et~al.}(2005)\citenamefont
  {Paulsson}, \citenamefont {Frederiksen},\ and\ \citenamefont
  {Brandbyge}}]{Paulsson:RapCom}%
  \BibitemOpen
  \bibfield  {author} {\bibinfo {author} {\bibfnamefont {M.}~\bibnamefont
  {Paulsson}}, \bibinfo {author} {\bibfnamefont {T.}~\bibnamefont
  {Frederiksen}}, \ and\ \bibinfo {author} {\bibfnamefont {M.}~\bibnamefont
  {Brandbyge}},\ }\Doi {10.1103/PhysRevB.72.201101} {\bibfield  {journal}
  {\bibinfo  {journal} {Phys. Rev. B},\ }\textbf {\bibinfo {volume} {72}},\
  \bibinfo {pages} {201101} (\bibinfo {year} {2005})}\BibitemShut {NoStop}%
\bibitem [{\citenamefont {Viljas}\ \emph {et~al.}(2005)\citenamefont {Viljas},
  \citenamefont {Cuevas}, \citenamefont {Pauly},\ and\ \citenamefont
  {H\"afner}}]{Viljas}%
  \BibitemOpen
  \bibfield  {author} {\bibinfo {author} {\bibfnamefont {J.~K.}\ \bibnamefont
  {Viljas}}, \bibinfo {author} {\bibfnamefont {J.~C.}\ \bibnamefont {Cuevas}},
  \bibinfo {author} {\bibfnamefont {F.}~\bibnamefont {Pauly}}, \ and\ \bibinfo
  {author} {\bibfnamefont {M.}~\bibnamefont {H\"afner}},\ }\Doi
  {10.1103/PhysRevB.72.245415} {\bibfield  {journal} {\bibinfo  {journal}
  {Phys. Rev. B},\ }\textbf {\bibinfo {volume} {72}},\ \bibinfo {pages}
  {245415} (\bibinfo {year} {2005})}\BibitemShut {NoStop}%
\bibitem [{\citenamefont {de~la Vega}\ \emph {et~al.}(2006)\citenamefont {de~la
  Vega}, \citenamefont {Mart\'\i{}n-Rodero}, \citenamefont {Agra\"\i{}t},\ and\
  \citenamefont {Yeyati}}]{delaVega}%
  \BibitemOpen
  \bibfield  {author} {\bibinfo {author} {\bibfnamefont {L.}~\bibnamefont
  {de~la Vega}}, \bibinfo {author} {\bibfnamefont {A.}~\bibnamefont
  {Mart\'\i{}n-Rodero}}, \bibinfo {author} {\bibfnamefont {N.}~\bibnamefont
  {Agra\"\i{}t}}, \ and\ \bibinfo {author} {\bibfnamefont {A.~L.}\ \bibnamefont
  {Yeyati}},\ }\Doi {10.1103/PhysRevB.73.075428} {\bibfield  {journal}
  {\bibinfo  {journal} {Phys. Rev. B},\ }\textbf {\bibinfo {volume} {73}},\
  \bibinfo {pages} {075428} (\bibinfo {year} {2006})}\BibitemShut {NoStop}%
\bibitem [{\citenamefont {Solomon}\ \emph {et~al.}(2006)\citenamefont
  {Solomon}, \citenamefont {Gagliardi}, \citenamefont {Pecchia}, \citenamefont
  {Frauenheim}, \citenamefont {Carlo}, \citenamefont {Reimers},\ and\
  \citenamefont {Hush}}]{Solomon}%
  \BibitemOpen
  \bibfield  {author} {\bibinfo {author} {\bibfnamefont {G.~C.}\ \bibnamefont
  {Solomon}}, \bibinfo {author} {\bibfnamefont {A.}~\bibnamefont {Gagliardi}},
  \bibinfo {author} {\bibfnamefont {A.}~\bibnamefont {Pecchia}}, \bibinfo
  {author} {\bibfnamefont {T.}~\bibnamefont {Frauenheim}}, \bibinfo {author}
  {\bibfnamefont {A.~D.}\ \bibnamefont {Carlo}}, \bibinfo {author}
  {\bibfnamefont {J.~R.}\ \bibnamefont {Reimers}}, \ and\ \bibinfo {author}
  {\bibfnamefont {N.~S.}\ \bibnamefont {Hush}},\ }\Doi {10.1063/1.2166362}
  {\bibfield  {journal} {\bibinfo  {journal} {J. Chem. Phys.},\ }\textbf
  {\bibinfo {volume} {124}},\ \bibinfo {eid} {094704} (\bibinfo {year}
  {2006})}\BibitemShut {NoStop}%
\bibitem [{\citenamefont {Sergueev}\ \emph {et~al.}(2007)\citenamefont
  {Sergueev}, \citenamefont {Demkov},\ and\ \citenamefont
  {Guo}}]{Sergueev:2007}%
  \BibitemOpen
  \bibfield  {author} {\bibinfo {author} {\bibfnamefont {N.}~\bibnamefont
  {Sergueev}}, \bibinfo {author} {\bibfnamefont {A.~A.}\ \bibnamefont
  {Demkov}}, \ and\ \bibinfo {author} {\bibfnamefont {H.}~\bibnamefont {Guo}},\
  }\Doi {10.1103/PhysRevB.75.233418} {\bibfield  {journal} {\bibinfo  {journal}
  {Phys. Rev. B},\ }\textbf {\bibinfo {volume} {75}},\ \bibinfo {pages}
  {233418} (\bibinfo {year} {2007})}\BibitemShut {NoStop}%
\bibitem [{\citenamefont {Frederiksen}\ \emph
  {et~al.}(2007){\natexlab{a}}\citenamefont {Frederiksen}, \citenamefont
  {Paulsson}, \citenamefont {Brandbyge},\ and\ \citenamefont
  {Jauho}}]{Frederiksen:PRB07}%
  \BibitemOpen
  \bibfield  {author} {\bibinfo {author} {\bibfnamefont {T.}~\bibnamefont
  {Frederiksen}}, \bibinfo {author} {\bibfnamefont {M.}~\bibnamefont
  {Paulsson}}, \bibinfo {author} {\bibfnamefont {M.}~\bibnamefont {Brandbyge}},
  \ and\ \bibinfo {author} {\bibfnamefont {A.-P.}\ \bibnamefont {Jauho}},\
  }\Doi {10.1103/PhysRevB.75.205413} {\bibfield  {journal} {\bibinfo  {journal}
  {Phys. Rev. B},\ }\textbf {\bibinfo {volume} {75}},\ \bibinfo {pages}
  {205413} (\bibinfo {year} {2007}{\natexlab{a}})}\BibitemShut {NoStop}%
\bibitem [{\citenamefont {Teobaldi}\ \emph {et~al.}(2007)\citenamefont
  {Teobaldi}, \citenamefont {Pe\~nalba}, \citenamefont {Arnau}, \citenamefont
  {Lorente},\ and\ \citenamefont {Hofer}}]{Teobaldi:2007}%
  \BibitemOpen
  \bibfield  {author} {\bibinfo {author} {\bibfnamefont {G.}~\bibnamefont
  {Teobaldi}}, \bibinfo {author} {\bibfnamefont {M.}~\bibnamefont {Pe\~nalba}},
  \bibinfo {author} {\bibfnamefont {A.}~\bibnamefont {Arnau}}, \bibinfo
  {author} {\bibfnamefont {N.}~\bibnamefont {Lorente}}, \ and\ \bibinfo
  {author} {\bibfnamefont {W.~A.}\ \bibnamefont {Hofer}},\ }\Doi
  {10.1103/PhysRevB.76.235407} {\bibfield  {journal} {\bibinfo  {journal}
  {Phys. Rev. B},\ }\textbf {\bibinfo {volume} {76}},\ \bibinfo {pages}
  {235407} (\bibinfo {year} {2007})}\BibitemShut {NoStop}%
\bibitem [{\citenamefont {Frederiksen}\ \emph
  {et~al.}(2007){\natexlab{b}}\citenamefont {Frederiksen}, \citenamefont
  {Lorente}, \citenamefont {Paulsson},\ and\ \citenamefont
  {Brandbyge}}]{Frederiksen:2007}%
  \BibitemOpen
  \bibfield  {author} {\bibinfo {author} {\bibfnamefont {T.}~\bibnamefont
  {Frederiksen}}, \bibinfo {author} {\bibfnamefont {N.}~\bibnamefont
  {Lorente}}, \bibinfo {author} {\bibfnamefont {M.}~\bibnamefont {Paulsson}}, \
  and\ \bibinfo {author} {\bibfnamefont {M.}~\bibnamefont {Brandbyge}},\ }\Doi
  {10.1103/PhysRevB.75.235441} {\bibfield  {journal} {\bibinfo  {journal}
  {Phys. Rev. B},\ }\textbf {\bibinfo {volume} {75}},\ \bibinfo {pages}
  {235441} (\bibinfo {year} {2007}{\natexlab{b}})}\BibitemShut {NoStop}%
\bibitem [{\citenamefont {Paulsson}\ \emph {et~al.}(2008)\citenamefont
  {Paulsson}, \citenamefont {Frederiksen}, \citenamefont {Ueba}, \citenamefont
  {Lorente},\ and\ \citenamefont {Brandbyge}}]{Paulsson:PRL100}%
  \BibitemOpen
  \bibfield  {author} {\bibinfo {author} {\bibfnamefont {M.}~\bibnamefont
  {Paulsson}}, \bibinfo {author} {\bibfnamefont {T.}~\bibnamefont
  {Frederiksen}}, \bibinfo {author} {\bibfnamefont {H.}~\bibnamefont {Ueba}},
  \bibinfo {author} {\bibfnamefont {N.}~\bibnamefont {Lorente}}, \ and\
  \bibinfo {author} {\bibfnamefont {M.}~\bibnamefont {Brandbyge}},\ }\Doi
  {10.1103/PhysRevLett.100.226604} {\bibfield  {journal} {\bibinfo  {journal}
  {Phys. Rev. Lett.},\ }\textbf {\bibinfo {volume} {100}},\ \bibinfo {pages}
  {226604} (\bibinfo {year} {2008})}\BibitemShut {NoStop}%
\bibitem [{\citenamefont {Kristensen}\ \emph {et~al.}(2009)\citenamefont
  {Kristensen}, \citenamefont {Paulsson}, \citenamefont {Thygesen},\ and\
  \citenamefont {Jacobsen}}]{Iben}%
  \BibitemOpen
  \bibfield  {author} {\bibinfo {author} {\bibfnamefont {I.~S.}\ \bibnamefont
  {Kristensen}}, \bibinfo {author} {\bibfnamefont {M.}~\bibnamefont
  {Paulsson}}, \bibinfo {author} {\bibfnamefont {K.~S.}\ \bibnamefont
  {Thygesen}}, \ and\ \bibinfo {author} {\bibfnamefont {K.~W.}\ \bibnamefont
  {Jacobsen}},\ }\Doi {10.1103/PhysRevB.79.235411} {\bibfield  {journal}
  {\bibinfo  {journal} {Phys. Rev. B},\ }\textbf {\bibinfo {volume} {79}},\
  \bibinfo {pages} {235411} (\bibinfo {year} {2009})}\BibitemShut {NoStop}%
\bibitem [{\citenamefont {Alducin}\ \emph {et~al.}(2010)\citenamefont
  {Alducin}, \citenamefont {S\'anchez-Portal}, \citenamefont {Arnau},\ and\
  \citenamefont {Lorente}}]{Alducin:PRL10}%
  \BibitemOpen
  \bibfield  {author} {\bibinfo {author} {\bibfnamefont {M.}~\bibnamefont
  {Alducin}}, \bibinfo {author} {\bibfnamefont {D.}~\bibnamefont
  {S\'anchez-Portal}}, \bibinfo {author} {\bibfnamefont {A.}~\bibnamefont
  {Arnau}}, \ and\ \bibinfo {author} {\bibfnamefont {N.}~\bibnamefont
  {Lorente}},\ }\Doi {10.1103/PhysRevLett.104.136101} {\bibfield  {journal}
  {\bibinfo  {journal} {Phys. Rev. Lett.},\ }\textbf {\bibinfo {volume}
  {104}},\ \bibinfo {pages} {136101} (\bibinfo {year} {2010})}\BibitemShut
  {NoStop}%
\bibitem [{\citenamefont {Fransson}\ \emph
  {et~al.}(2010){\natexlab{a}}\citenamefont {Fransson}, \citenamefont
  {Manoharan},\ and\ \citenamefont {Balatsky}}]{Fransson:NL10}%
  \BibitemOpen
  \bibfield  {author} {\bibinfo {author} {\bibfnamefont {J.}~\bibnamefont
  {Fransson}}, \bibinfo {author} {\bibfnamefont {H.~C.}\ \bibnamefont
  {Manoharan}}, \ and\ \bibinfo {author} {\bibfnamefont {A.~V.}\ \bibnamefont
  {Balatsky}},\ }\Doi {10.1021/nl903991a} {\bibfield  {journal} {\bibinfo
  {journal} {Nano Lett.},\ }\textbf {\bibinfo {volume} {10}},\ \bibinfo {pages}
  {1600} (\bibinfo {year} {2010}{\natexlab{a}})}\BibitemShut {NoStop}%
\bibitem [{\citenamefont {Monturet}\ \emph {et~al.}(2010)\citenamefont
  {Monturet}, \citenamefont {Alducin},\ and\ \citenamefont
  {Lorente}}]{Monturet:2010}%
  \BibitemOpen
  \bibfield  {author} {\bibinfo {author} {\bibfnamefont {S.}~\bibnamefont
  {Monturet}}, \bibinfo {author} {\bibfnamefont {M.}~\bibnamefont {Alducin}}, \
  and\ \bibinfo {author} {\bibfnamefont {N.}~\bibnamefont {Lorente}},\ }\Doi
  {10.1103/PhysRevB.82.085447} {\bibfield  {journal} {\bibinfo  {journal}
  {Phys. Rev. B},\ }\textbf {\bibinfo {volume} {82}},\ \bibinfo {pages}
  {085447} (\bibinfo {year} {2010})}\BibitemShut {NoStop}%
\bibitem [{\citenamefont {Patton}(2010)}]{Patton}%
  \BibitemOpen
  \bibfield  {author} {\bibinfo {author} {\bibfnamefont {K.~R.}\ \bibnamefont
  {Patton}},\ }\href {http://arxiv.org/abs/1007.1238v1} {\bibfield  {journal}
  {\bibinfo  {journal} {arXiv:1007.1238v1}} (\bibinfo {year} {2010})},\
  \bibinfo {note} {(unpublished)}\BibitemShut {NoStop}%
\bibitem [{\citenamefont {Beenakker}\ and\ \citenamefont
  {Sch\"{o}nenberger}(2003)}]{Beenakker}%
  \BibitemOpen
  \bibfield  {author} {\bibinfo {author} {\bibfnamefont {C.}~\bibnamefont
  {Beenakker}}\ and\ \bibinfo {author} {\bibfnamefont {C.}~\bibnamefont
  {Sch\"{o}nenberger}},\ }\Doi {10.1063/1.1583532} {\bibfield  {journal}
  {\bibinfo  {journal} {Physics Today},\ }\textbf {\bibinfo {volume} {56}},\
  \bibinfo {pages} {37} (\bibinfo {year} {2003})}\BibitemShut {NoStop}%
\bibitem [{\citenamefont {Blanter}\ and\ \citenamefont
  {B\"uttiker}(2000)}]{Blanter}%
  \BibitemOpen
  \bibfield  {author} {\bibinfo {author} {\bibfnamefont {Y.~M.}\ \bibnamefont
  {Blanter}}\ and\ \bibinfo {author} {\bibfnamefont {M.}~\bibnamefont
  {B\"uttiker}},\ }\Doi {DOI: 10.1016/S0370-1573(99)00123-4} {\bibfield
  {journal} {\bibinfo  {journal} {Phys. Rep.},\ }\textbf {\bibinfo {volume}
  {336}},\ \bibinfo {pages} {1 } (\bibinfo {year} {2000})}\BibitemShut
  {NoStop}%
\bibitem [{\citenamefont {Nazarov}(1999)}]{Nazarov:AnnPhys}%
  \BibitemOpen
  \bibfield  {author} {\bibinfo {author} {\bibfnamefont {Y.~V.}\ \bibnamefont
  {Nazarov}},\ }\href@noop {} {\bibfield  {journal} {\bibinfo  {journal} {Ann.
  Phys.},\ }\textbf {\bibinfo {volume} {8}},\ \bibinfo {pages} {SI-193} (\bibinfo
  {year} {1999})}\BibitemShut {NoStop}%
\bibitem [{\citenamefont {Nazarov}(2003)}]{Nazarov:book}%
  \BibitemOpen
  \bibinfo {editor} {\bibfnamefont {Y.~V.}\ \bibnamefont {Nazarov}},\ ed.,\
  \href@noop {} {\emph {\bibinfo {title} {Quantum Noise in Mesoscopic
  Physics}}}\ (\bibinfo  {publisher} {Springer},\ \bibinfo {address} {Berlin},\
  \bibinfo {year} {2003})\BibitemShut {NoStop}%
\bibitem [{\citenamefont {van~den Brom}\ and\ \citenamefont {van
  Ruitenbeek}(1999)}]{VandenBrom}%
  \BibitemOpen
  \bibfield  {author} {\bibinfo {author} {\bibfnamefont {H.~E.}\ \bibnamefont
  {van~den Brom}}\ and\ \bibinfo {author} {\bibfnamefont {J.~M.}\ \bibnamefont
  {van Ruitenbeek}},\ }\Doi {10.1103/PhysRevLett.82.1526} {\bibfield  {journal}
  {\bibinfo  {journal} {Phys. Rev. Lett.},\ }\textbf {\bibinfo {volume} {82}},\
  \bibinfo {pages} {1526} (\bibinfo {year} {1999})}\BibitemShut {NoStop}%
\bibitem [{\citenamefont {Djukic}\ and\ \citenamefont {van
  Ruitenbeek}(2006)}]{Djukic}%
  \BibitemOpen
  \bibfield  {author} {\bibinfo {author} {\bibfnamefont {D.}~\bibnamefont
  {Djukic}}\ and\ \bibinfo {author} {\bibfnamefont {J.~M.}\ \bibnamefont {van
  Ruitenbeek}},\ }\Doi {10.1021/nl060116e} {\bibfield  {journal} {\bibinfo
  {journal} {Nano Lett.},\ }\textbf {\bibinfo {volume} {6}},\ \bibinfo {pages}
  {789} (\bibinfo {year} {2006})}\BibitemShut {NoStop}%
\bibitem [{\citenamefont {Kiguchi}\ \emph {et~al.}(2008)\citenamefont
  {Kiguchi}, \citenamefont {Tal}, \citenamefont {Wohlthat}, \citenamefont
  {Pauly}, \citenamefont {Krieger}, \citenamefont {Djukic}, \citenamefont
  {Cuevas},\ and\ \citenamefont {van Ruitenbeek}}]{Kiguchi}%
  \BibitemOpen
  \bibfield  {author} {\bibinfo {author} {\bibfnamefont {M.}~\bibnamefont
  {Kiguchi}}, \bibinfo {author} {\bibfnamefont {O.}~\bibnamefont {Tal}},
  \bibinfo {author} {\bibfnamefont {S.}~\bibnamefont {Wohlthat}}, \bibinfo
  {author} {\bibfnamefont {F.}~\bibnamefont {Pauly}}, \bibinfo {author}
  {\bibfnamefont {M.}~\bibnamefont {Krieger}}, \bibinfo {author} {\bibfnamefont
  {D.}~\bibnamefont {Djukic}}, \bibinfo {author} {\bibfnamefont {J.~C.}\
  \bibnamefont {Cuevas}}, \ and\ \bibinfo {author} {\bibfnamefont {J.~M.}\
  \bibnamefont {van Ruitenbeek}},\ }\Doi {10.1103/PhysRevLett.101.046801}
  {\bibfield  {journal} {\bibinfo  {journal} {Phys. Rev. Lett.},\ }\textbf
  {\bibinfo {volume} {101}},\ \bibinfo {pages} {046801} (\bibinfo {year}
  {2008})}\BibitemShut {NoStop}%
\bibitem [{Note1()}]{Note1}%
  \BibitemOpen
  \bibinfo {note} {J.~M.~van Ruitenbeek, private communication.}\BibitemShut
  {Stop}%
\bibitem [{\citenamefont {Mitra}\ \emph {et~al.}(2004)\citenamefont {Mitra},
  \citenamefont {Aleiner},\ and\ \citenamefont {Millis}}]{Mitra}%
  \BibitemOpen
  \bibfield  {author} {\bibinfo {author} {\bibfnamefont {A.}~\bibnamefont
  {Mitra}}, \bibinfo {author} {\bibfnamefont {I.}~\bibnamefont {Aleiner}}, \
  and\ \bibinfo {author} {\bibfnamefont {A.~J.}\ \bibnamefont {Millis}},\ }\Doi
  {10.1103/PhysRevB.69.245302} {\bibfield  {journal} {\bibinfo  {journal}
  {Phys. Rev. B},\ }\textbf {\bibinfo {volume} {69}},\ \bibinfo {pages}
  {245302} (\bibinfo {year} {2004})}\BibitemShut {NoStop}%
\bibitem [{\citenamefont {Koch}\ and\ \citenamefont {von
  Oppen}(2005)}]{KochPRL94}%
  \BibitemOpen
  \bibfield  {author} {\bibinfo {author} {\bibfnamefont {J.}~\bibnamefont
  {Koch}}\ and\ \bibinfo {author} {\bibfnamefont {F.}~\bibnamefont {von
  Oppen}},\ }\Doi {10.1103/PhysRevLett.94.206804} {\bibfield  {journal}
  {\bibinfo  {journal} {Phys. Rev. Lett.},\ }\textbf {\bibinfo {volume} {94}},\
  \bibinfo {pages} {206804} (\bibinfo {year} {2005})}\BibitemShut {NoStop}%
\bibitem [{\citenamefont {Koch}\ \emph {et~al.}(2005)\citenamefont {Koch},
  \citenamefont {Raikh},\ and\ \citenamefont {von Oppen}}]{KochPRL95}%
  \BibitemOpen
  \bibfield  {author} {\bibinfo {author} {\bibfnamefont {J.}~\bibnamefont
  {Koch}}, \bibinfo {author} {\bibfnamefont {M.~E.}\ \bibnamefont {Raikh}}, \
  and\ \bibinfo {author} {\bibfnamefont {F.}~\bibnamefont {von Oppen}},\ }\Doi
  {10.1103/PhysRevLett.95.056801} {\bibfield  {journal} {\bibinfo  {journal}
  {Phys. Rev. Lett.},\ }\textbf {\bibinfo {volume} {95}},\ \bibinfo {pages}
  {056801} (\bibinfo {year} {2005})}\BibitemShut {NoStop}%
\bibitem [{\citenamefont {Zhu}\ and\ \citenamefont {Balatsky}(2003)}]{Zhu}%
  \BibitemOpen
  \bibfield  {author} {\bibinfo {author} {\bibfnamefont {J.-X.}\ \bibnamefont
  {Zhu}}\ and\ \bibinfo {author} {\bibfnamefont {A.~V.}\ \bibnamefont
  {Balatsky}},\ }\Doi {10.1103/PhysRevB.67.165326} {\bibfield  {journal}
  {\bibinfo  {journal} {Phys. Rev. B},\ }\textbf {\bibinfo {volume} {67}},\
  \bibinfo {pages} {165326} (\bibinfo {year} {2003})}\BibitemShut {NoStop}%
\bibitem [{\citenamefont {Galperin}\ \emph {et~al.}(2006)\citenamefont
  {Galperin}, \citenamefont {Nitzan},\ and\ \citenamefont
  {Ratner}}]{Galperin:PRB06}%
  \BibitemOpen
  \bibfield  {author} {\bibinfo {author} {\bibfnamefont {M.}~\bibnamefont
  {Galperin}}, \bibinfo {author} {\bibfnamefont {A.}~\bibnamefont {Nitzan}}, \
  and\ \bibinfo {author} {\bibfnamefont {M.~A.}\ \bibnamefont {Ratner}},\ }\Doi
  {10.1103/PhysRevB.74.075326} {\bibfield  {journal} {\bibinfo  {journal}
  {Phys. Rev. B},\ }\textbf {\bibinfo {volume} {74}},\ \bibinfo {pages}
  {075326} (\bibinfo {year} {2006})}\BibitemShut {NoStop}%
\bibitem [{\citenamefont {Levitov}\ and\ \citenamefont
  {Lesovik}(1993)}]{Lesovik}%
  \BibitemOpen
  \bibfield  {author} {\bibinfo {author} {\bibfnamefont {L.~S.}\ \bibnamefont
  {Levitov}}\ and\ \bibinfo {author} {\bibfnamefont {G.~B.}\ \bibnamefont
  {Lesovik}},\ }\href@noop {} {\bibfield  {journal} {\bibinfo  {journal} {JETP
  Lett.},\ }\textbf {\bibinfo {volume} {58}},\ \bibinfo {pages} {230} (\bibinfo
  {year} {1993})}\BibitemShut {NoStop}%
\bibitem [{\citenamefont {Levitov}\ \emph {et~al.}(1996)\citenamefont
  {Levitov}, \citenamefont {Lee},\ and\ \citenamefont {Lesovik}}]{LLL:JMP}%
  \BibitemOpen
  \bibfield  {author} {\bibinfo {author} {\bibfnamefont {L.~S.}\ \bibnamefont
  {Levitov}}, \bibinfo {author} {\bibfnamefont {H.~W.}\ \bibnamefont {Lee}}, \
  and\ \bibinfo {author} {\bibfnamefont {G.~B.}\ \bibnamefont {Lesovik}},\
  }\href@noop {} {\bibfield  {journal} {\bibinfo  {journal} {J. Math. Phys.},\
  }\textbf {\bibinfo {volume} {37}},\ \bibinfo {pages} {4845} (\bibinfo {year}
  {1996})}\BibitemShut {NoStop}%
\bibitem [{\citenamefont {Schmidt}\ and\ \citenamefont
  {Komnik}(2009)}]{Schmidt}%
  \BibitemOpen
  \bibfield  {author} {\bibinfo {author} {\bibfnamefont {T.~L.}\ \bibnamefont
  {Schmidt}}\ and\ \bibinfo {author} {\bibfnamefont {A.}~\bibnamefont
  {Komnik}},\ }\Doi {10.1103/PhysRevB.80.041307} {\bibfield  {journal}
  {\bibinfo  {journal} {Phys. Rev. B},\ }\textbf {\bibinfo {volume} {80}},\
  \bibinfo {pages} {041307} (\bibinfo {year} {2009})}\BibitemShut {NoStop}%
\bibitem [{\citenamefont {Avriller}\ and\ \citenamefont
  {Levy~Yeyati}(2009)}]{Avriller}%
  \BibitemOpen
  \bibfield  {author} {\bibinfo {author} {\bibfnamefont {R.}~\bibnamefont
  {Avriller}}\ and\ \bibinfo {author} {\bibfnamefont {A.}~\bibnamefont
  {Levy~Yeyati}},\ }\Doi {10.1103/PhysRevB.80.041309} {\bibfield  {journal}
  {\bibinfo  {journal} {Phys. Rev. B},\ }\textbf {\bibinfo {volume} {80}},\
  \bibinfo {pages} {041309} (\bibinfo {year} {2009})}\BibitemShut {NoStop}%
\bibitem [{\citenamefont {Haupt}\ \emph {et~al.}(2009)\citenamefont {Haupt},
  \citenamefont {Novotn\'y},\ and\ \citenamefont {Belzig}}]{Haupt}%
  \BibitemOpen
  \bibfield  {author} {\bibinfo {author} {\bibfnamefont {F.}~\bibnamefont
  {Haupt}}, \bibinfo {author} {\bibfnamefont {T.}~\bibnamefont {Novotn\'y}}, \
  and\ \bibinfo {author} {\bibfnamefont {W.}~\bibnamefont {Belzig}},\ }\Doi
  {10.1103/PhysRevLett.103.136601} {\bibfield  {journal} {\bibinfo  {journal}
  {Phys. Rev. Lett.},\ }\textbf {\bibinfo {volume} {103}},\ \bibinfo {pages}
  {136601} (\bibinfo {year} {2009})}\BibitemShut {NoStop}%
\bibitem [{Note2()}]{Note2}%
  \BibitemOpen
  \bibinfo {note} {Based on a spin-less model, our results need to be
  multiplied by a factor of 2 when compared with works where spin degeneracy is
  explicitly taken into account.}\BibitemShut {Stop}%
\bibitem [{\citenamefont {Levitov}\ and\ \citenamefont
  {Reznikov}(2004)}]{Levitov}%
  \BibitemOpen
  \bibfield  {author} {\bibinfo {author} {\bibfnamefont {L.~S.}\ \bibnamefont
  {Levitov}}\ and\ \bibinfo {author} {\bibfnamefont {M.}~\bibnamefont
  {Reznikov}},\ }\Doi {10.1103/PhysRevB.70.115305} {\bibfield  {journal}
  {\bibinfo  {journal} {Phys. Rev. B},\ }\textbf {\bibinfo {volume} {70}},\
  \bibinfo {pages} {115305} (\bibinfo {year} {2004})}\BibitemShut {NoStop}%
\bibitem [{\citenamefont {Gogolin}\ and\ \citenamefont
  {Komnik}(2006)}]{GogolinKomnik}%
  \BibitemOpen
  \bibfield  {author} {\bibinfo {author} {\bibfnamefont {A.~O.}\ \bibnamefont
  {Gogolin}}\ and\ \bibinfo {author} {\bibfnamefont {A.}~\bibnamefont
  {Komnik}},\ }\Doi {10.1103/PhysRevB.73.195301} {\bibfield  {journal}
  {\bibinfo  {journal} {Phys. Rev. B},\ }\textbf {\bibinfo {volume} {73}},\
  \bibinfo {pages} {195301} (\bibinfo {year} {2006})}\BibitemShut {NoStop}%
\bibitem [{\citenamefont {Souza}\ \emph {et~al.}(2008)\citenamefont {Souza},
  \citenamefont {Jauho},\ and\ \citenamefont {Egues}}]{Souza}%
  \BibitemOpen
  \bibfield  {author} {\bibinfo {author} {\bibfnamefont {F.~M.}\ \bibnamefont
  {Souza}}, \bibinfo {author} {\bibfnamefont {A.~P.}\ \bibnamefont {Jauho}}, \
  and\ \bibinfo {author} {\bibfnamefont {J.~C.}\ \bibnamefont {Egues}},\ }\Doi
  {10.1103/PhysRevB.78.155303} {\bibfield  {journal} {\bibinfo  {journal}
  {Phys. Rev. B},\ }\textbf {\bibinfo {volume} {78}},\ \bibinfo {pages}
  {155303} (\bibinfo {year} {2008})}\BibitemShut {NoStop}%
\bibitem [{\citenamefont {Meir}\ and\ \citenamefont {Wingreen}(1992)}]{Meir}%
  \BibitemOpen
  \bibfield  {author} {\bibinfo {author} {\bibfnamefont {Y.}~\bibnamefont
  {Meir}}\ and\ \bibinfo {author} {\bibfnamefont {N.~S.}\ \bibnamefont
  {Wingreen}},\ }\Doi {10.1103/PhysRevLett.68.2512} {\bibfield  {journal}
  {\bibinfo  {journal} {Phys. Rev. Lett.},\ }\textbf {\bibinfo {volume} {68}},\
  \bibinfo {pages} {2512} (\bibinfo {year} {1992})}\BibitemShut {NoStop}%
\bibitem [{\citenamefont {B\"uttiker}(1992)}]{Buettiker}%
  \BibitemOpen
  \bibfield  {author} {\bibinfo {author} {\bibfnamefont {M.}~\bibnamefont
  {B\"uttiker}},\ }\Doi {10.1103/PhysRevB.46.12485} {\bibfield  {journal}
  {\bibinfo  {journal} {Phys. Rev. B},\ }\textbf {\bibinfo {volume} {46}},\
  \bibinfo {pages} {12485} (\bibinfo {year} {1992})}\BibitemShut {NoStop}%
\bibitem [{\citenamefont {Paulsson}\ and\ \citenamefont
  {Brandbyge}(2007)}]{PaulssonBrandbyge}%
  \BibitemOpen
  \bibfield  {author} {\bibinfo {author} {\bibfnamefont {M.}~\bibnamefont
  {Paulsson}}\ and\ \bibinfo {author} {\bibfnamefont {M.}~\bibnamefont
  {Brandbyge}},\ }\Doi {10.1103/PhysRevB.76.115117} {\bibfield  {journal}
  {\bibinfo  {journal} {Phys. Rev. B},\ }\textbf {\bibinfo {volume} {76}},\
  \bibinfo {pages} {115117} (\bibinfo {year} {2007})}\BibitemShut {NoStop}%
\bibitem [{\citenamefont {Fransson}\ and\ \citenamefont
  {Galperin}(2010)}]{Fransson}%
  \BibitemOpen
  \bibfield  {author} {\bibinfo {author} {\bibfnamefont {J.}~\bibnamefont
  {Fransson}}\ and\ \bibinfo {author} {\bibfnamefont {M.}~\bibnamefont
  {Galperin}},\ }\Doi {10.1103/PhysRevB.81.075311} {\bibfield  {journal}
  {\bibinfo  {journal} {Phys. Rev. B},\ }\textbf {\bibinfo {volume} {81}},\
  \bibinfo {pages} {075311} (\bibinfo {year} {2010})}\BibitemShut {NoStop}%
\bibitem [{Note3()}]{Note3}%
  \BibitemOpen
  \bibinfo {note} {Unlike $ \protect \ensuremath {\protect \mathbf {n}}_{e}$,
  the evaluation of $\protect \ensuremath {\protect \mathbf {n}}_{\nu }'$
  involves only integrands with compact support.}\BibitemShut {Stop}%
\bibitem [{\citenamefont {Hershfield}(1992)}]{Hershfield}%
  \BibitemOpen
  \bibfield  {author} {\bibinfo {author} {\bibfnamefont {S.}~\bibnamefont
  {Hershfield}},\ }\Doi {10.1103/PhysRevB.46.7061} {\bibfield  {journal}
  {\bibinfo  {journal} {Phys. Rev. B},\ }\textbf {\bibinfo {volume} {46}},\
  \bibinfo {pages} {7061} (\bibinfo {year} {1992})}\BibitemShut {NoStop}%
\bibitem [{Note4()}]{Note4}%
  \BibitemOpen
  \bibinfo {note} {See associated \protect \texttt {Mathematica} notebook for
  the complete expression for $S_{F}^{\protect \rm (mf/vc)}$ at finite
  temperature.}\BibitemShut {Stop}%
\bibitem [{\citenamefont {Entin-Wohlman}\ \emph {et~al.}(2009)\citenamefont
  {Entin-Wohlman}, \citenamefont {Imry},\ and\ \citenamefont
  {Aharony}}]{Entin-Wohlman}%
  \BibitemOpen
  \bibfield  {author} {\bibinfo {author} {\bibfnamefont {O.}~\bibnamefont
  {Entin-Wohlman}}, \bibinfo {author} {\bibfnamefont {Y.}~\bibnamefont {Imry}},
  \ and\ \bibinfo {author} {\bibfnamefont {A.}~\bibnamefont {Aharony}},\ }\Doi
  {10.1103/PhysRevB.80.035417} {\bibfield  {journal} {\bibinfo  {journal}
  {Phys. Rev. B},\ }\textbf {\bibinfo {volume} {80}},\ \bibinfo {pages}
  {035417} (\bibinfo {year} {2009})}\BibitemShut {NoStop}%
\bibitem [{\citenamefont {Egger}\ and\ \citenamefont {Gogolin}(2008)}]{Egger}%
  \BibitemOpen
  \bibfield  {author} {\bibinfo {author} {\bibfnamefont {R.}~\bibnamefont
  {Egger}}\ and\ \bibinfo {author} {\bibfnamefont {A.~O.}\ \bibnamefont
  {Gogolin}},\ }\Doi {10.1103/PhysRevB.77.113405} {\bibfield  {journal}
  {\bibinfo  {journal} {Phys. Rev. B},\ }\textbf {\bibinfo {volume} {77}},\
  \bibinfo {pages} {113405} (\bibinfo {year} {2008})}\BibitemShut {NoStop}%
\bibitem [{\citenamefont {Engelund}\ \emph {et~al.}(2009)\citenamefont
  {Engelund}, \citenamefont {Brandbyge},\ and\ \citenamefont
  {Jauho}}]{Engelund}%
  \BibitemOpen
  \bibfield  {author} {\bibinfo {author} {\bibfnamefont {M.}~\bibnamefont
  {Engelund}}, \bibinfo {author} {\bibfnamefont {M.}~\bibnamefont {Brandbyge}},
  \ and\ \bibinfo {author} {\bibfnamefont {A.~P.}\ \bibnamefont {Jauho}},\
  }\Doi {10.1103/PhysRevB.80.045427} {\bibfield  {journal} {\bibinfo  {journal}
  {Phys. Rev. B},\ }\textbf {\bibinfo {volume} {80}},\ \bibinfo {pages}
  {045427} (\bibinfo {year} {2009})}\BibitemShut {NoStop}%
\bibitem [{\citenamefont {Asai}(2008)}]{Asai}%
  \BibitemOpen
  \bibfield  {author} {\bibinfo {author} {\bibfnamefont {Y.}~\bibnamefont
  {Asai}},\ }\Doi {10.1103/PhysRevB.78.045434} {\bibfield  {journal} {\bibinfo
  {journal} {Phys. Rev. B},\ }\textbf {\bibinfo {volume} {78}},\ \bibinfo
  {pages} {045434} (\bibinfo {year} {2008})}\BibitemShut {NoStop}%
\bibitem [{\citenamefont {Ryndyk}\ and\ \citenamefont
  {Cuniberti}(2007)}]{Ryndyk}%
  \BibitemOpen
  \bibfield  {author} {\bibinfo {author} {\bibfnamefont {D.~A.}\ \bibnamefont
  {Ryndyk}}\ and\ \bibinfo {author} {\bibfnamefont {G.}~\bibnamefont
  {Cuniberti}},\ }\Doi {10.1103/PhysRevB.76.155430} {\bibfield  {journal}
  {\bibinfo  {journal} {Phys. Rev. B},\ }\textbf {\bibinfo {volume} {76}},\
  \bibinfo {pages} {155430} (\bibinfo {year} {2007})}\BibitemShut {NoStop}%
\bibitem [{Note5()}]{Note5}%
  \BibitemOpen
  \bibinfo {note} {Its generalization to multilevel case is straightforward and
  follows exactly the lines of Refs.~\protect \rev@citealpnum
  {Frederiksen:PRL04,Frederiksen:PRB07}.}\BibitemShut {Stop}%
\bibitem [{\citenamefont {Urban}\ \emph {et~al.}(2010)\citenamefont {Urban},
  \citenamefont {Avriller},\ and\ \citenamefont {Levy~Yeyati}}]{Urban}%
  \BibitemOpen
  \bibfield  {author} {\bibinfo {author} {\bibfnamefont {D.~F.}\ \bibnamefont
  {Urban}}, \bibinfo {author} {\bibfnamefont {R.}~\bibnamefont {Avriller}}, \
  and\ \bibinfo {author} {\bibfnamefont {A.}~\bibnamefont {Levy~Yeyati}},\
  }\Doi {10.1103/PhysRevB.82.121414} {\bibfield  {journal} {\bibinfo  {journal}
  {Phys. Rev. B},\ }\textbf {\bibinfo {volume} {82}},\ \bibinfo {pages}
  {121414} (\bibinfo {year} {2010})}\BibitemShut {NoStop}%
\bibitem [{\citenamefont {Jouravlev}(2005)}]{Jouravlev}%
  \BibitemOpen
  \bibfield  {author} {\bibinfo {author} {\bibfnamefont {O.~N.}\ \bibnamefont
  {Jouravlev}},\ }\emph {\bibinfo {title} {Noise and Spin in Nanostructures}},\
  \href@noop {} {Ph.D. thesis},\ \bibinfo  {school} {TU Delft}, \bibinfo
  {address} {Netherlands} (\bibinfo {year} {2005})\BibitemShut {NoStop}%
\bibitem [{\citenamefont
  {Fern\'andez-Rossier}(2009)}]{Fernandez-Rossier:PRL09}%
  \BibitemOpen
  \bibfield  {author} {\bibinfo {author} {\bibfnamefont {J.}~\bibnamefont
  {Fern\'andez-Rossier}},\ }\Doi {10.1103/PhysRevLett.102.256802} {\bibfield
  {journal} {\bibinfo  {journal} {Phys. Rev. Lett.},\ }\textbf {\bibinfo
  {volume} {102}},\ \bibinfo {pages} {256802} (\bibinfo {year}
  {2009})}\BibitemShut {NoStop}%
\bibitem [{\citenamefont {Persson}(2009)}]{Persson:PRL09}%
  \BibitemOpen
  \bibfield  {author} {\bibinfo {author} {\bibfnamefont {M.}~\bibnamefont
  {Persson}},\ }\Doi {10.1103/PhysRevLett.103.050801} {\bibfield  {journal}
  {\bibinfo  {journal} {Phys. Rev. Lett.},\ }\textbf {\bibinfo {volume}
  {103}},\ \bibinfo {pages} {050801} (\bibinfo {year} {2009})}\BibitemShut
  {NoStop}%
\bibitem [{\citenamefont {Fransson}(2009)}]{Fransson:NL09}%
  \BibitemOpen
  \bibfield  {author} {\bibinfo {author} {\bibfnamefont {J.}~\bibnamefont
  {Fransson}},\ }\Doi {10.1021/nl901066a} {\bibfield  {journal} {\bibinfo
  {journal} {Nano Lett.},\ }\textbf {\bibinfo {volume} {9}},\ \bibinfo {pages}
  {2414} (\bibinfo {year} {2009})}\BibitemShut {NoStop}%
\bibitem [{\citenamefont {Lorente}\ and\ \citenamefont
  {Gauyacq}(2009)}]{Lorente:PRL09}%
  \BibitemOpen
  \bibfield  {author} {\bibinfo {author} {\bibfnamefont {N.}~\bibnamefont
  {Lorente}}\ and\ \bibinfo {author} {\bibfnamefont {J.-P.}\ \bibnamefont
  {Gauyacq}},\ }\Doi {10.1103/PhysRevLett.103.176601} {\bibfield  {journal}
  {\bibinfo  {journal} {Phys. Rev. Lett.},\ }\textbf {\bibinfo {volume}
  {103}},\ \bibinfo {pages} {176601} (\bibinfo {year} {2009})}\BibitemShut
  {NoStop}%
\bibitem [{\citenamefont {Gauyacq}\ \emph {et~al.}(2010)\citenamefont
  {Gauyacq}, \citenamefont {Novaes},\ and\ \citenamefont
  {Lorente}}]{Gauyacq:PRB10}%
  \BibitemOpen
  \bibfield  {author} {\bibinfo {author} {\bibfnamefont {J.-P.}\ \bibnamefont
  {Gauyacq}}, \bibinfo {author} {\bibfnamefont {F.~D.}\ \bibnamefont {Novaes}},
  \ and\ \bibinfo {author} {\bibfnamefont {N.}~\bibnamefont {Lorente}},\ }\Doi
  {10.1103/PhysRevB.81.165423} {\bibfield  {journal} {\bibinfo  {journal}
  {Phys. Rev. B},\ }\textbf {\bibinfo {volume} {81}},\ \bibinfo {pages}
  {165423} (\bibinfo {year} {2010})}\BibitemShut {NoStop}%
\bibitem [{\citenamefont {Novaes}\ \emph {et~al.}(2010)\citenamefont {Novaes},
  \citenamefont {Lorente},\ and\ \citenamefont {Gauyacq}}]{Novaes:preprint}%
  \BibitemOpen
  \bibfield  {author} {\bibinfo {author} {\bibfnamefont {F.~D.}\ \bibnamefont
  {Novaes}}, \bibinfo {author} {\bibfnamefont {N.}~\bibnamefont {Lorente}}, \
  and\ \bibinfo {author} {\bibfnamefont {J.-P.}\ \bibnamefont {Gauyacq}},\
  }\Doi {10.1103/PhysRevB.82.155401} {\bibfield  {journal} {\bibinfo  {journal}
  {Phys. Rev. B},\ }\textbf {\bibinfo {volume} {82}},\ \bibinfo {pages}
  {155401} (\bibinfo {year} {2010})}\BibitemShut {NoStop}%
\bibitem [{\citenamefont {Delgado}\ \emph {et~al.}(2010)\citenamefont
  {Delgado}, \citenamefont {Palacios},\ and\ \citenamefont
  {Fern\'andez-Rossier}}]{Delgado:PRL10}%
  \BibitemOpen
  \bibfield  {author} {\bibinfo {author} {\bibfnamefont {F.}~\bibnamefont
  {Delgado}}, \bibinfo {author} {\bibfnamefont {J.~J.}\ \bibnamefont
  {Palacios}}, \ and\ \bibinfo {author} {\bibfnamefont {J.}~\bibnamefont
  {Fern\'andez-Rossier}},\ }\Doi {10.1103/PhysRevLett.104.026601} {\bibfield
  {journal} {\bibinfo  {journal} {Phys. Rev. Lett.},\ }\textbf {\bibinfo
  {volume} {104}},\ \bibinfo {pages} {026601} (\bibinfo {year}
  {2010})}\BibitemShut {NoStop}%
\bibitem [{\citenamefont {Delgado}\ and\ \citenamefont
  {Fern{\'a}ndez-Rossier}(2010)}]{Delgado:preprint}%
  \BibitemOpen
  \bibfield  {author} {\bibinfo {author} {\bibfnamefont {F.}~\bibnamefont
  {Delgado}}\ and\ \bibinfo {author} {\bibfnamefont {J.}~\bibnamefont
  {Fern{\'a}ndez-Rossier}},\ }\href {http://arxiv.org/abs/1006.5608v1}
  {\bibfield  {journal} {\bibinfo  {journal} {arXiv:1006.5608v1}} (\bibinfo
  {year} {2010})}\BibitemShut {NoStop}%
\bibitem [{\citenamefont {Fransson}\ \emph
  {et~al.}(2010){\natexlab{b}}\citenamefont {Fransson}, \citenamefont
  {Eriksson},\ and\ \citenamefont {Balatsky}}]{Balatsky}%
  \BibitemOpen
  \bibfield  {author} {\bibinfo {author} {\bibfnamefont {J.}~\bibnamefont
  {Fransson}}, \bibinfo {author} {\bibfnamefont {O.}~\bibnamefont {Eriksson}},
  \ and\ \bibinfo {author} {\bibfnamefont {A.~V.}\ \bibnamefont {Balatsky}},\
  }\Doi {10.1103/PhysRevB.81.115454} {\bibfield  {journal} {\bibinfo  {journal}
  {Phys. Rev. B},\ }\textbf {\bibinfo {volume} {81}},\ \bibinfo {pages}
  {115454} (\bibinfo {year} {2010}{\natexlab{b}})}\BibitemShut {NoStop}%
\bibitem [{\citenamefont {Elste}\ and\ \citenamefont {Timm}(2010)}]{Timm}%
  \BibitemOpen
  \bibfield  {author} {\bibinfo {author} {\bibfnamefont {F.}~\bibnamefont
  {Elste}}\ and\ \bibinfo {author} {\bibfnamefont {C.}~\bibnamefont {Timm}},\
  }\Doi {10.1103/PhysRevB.81.024421} {\bibfield  {journal} {\bibinfo  {journal}
  {Phys. Rev. B},\ }\textbf {\bibinfo {volume} {81}},\ \bibinfo {pages}
  {024421} (\bibinfo {year} {2010})}\BibitemShut {NoStop}%
\bibitem [{Note6()}]{Note6}%
  \BibitemOpen
  \bibinfo {note} {Apart from an extra factor of 2 in Eqs. (30), (31) of
  Ref.~\protect \rev@citealpnum {Souza} stemming from their different
  definition of the zero-frequency noise.}\BibitemShut {Stop}%
\bibitem [{\citenamefont {Haug}\ and\ \citenamefont
  {Jauho}(2008)}]{Jauho:book}%
  \BibitemOpen
  \bibfield  {author} {\bibinfo {author} {\bibfnamefont {H.}~\bibnamefont
  {Haug}}\ and\ \bibinfo {author} {\bibfnamefont {A.-P.}\ \bibnamefont
  {Jauho}},\ }\href@noop {} {\emph {\bibinfo {title} {Quantum Kinetics in
  Transport and Optics of Semiconductors}}},\ \bibinfo {edition} {2nd}\ ed.\
  (\bibinfo  {publisher} {Springer},\ \bibinfo {address} {Berlin},\ \bibinfo
  {year} {2008})\BibitemShut {NoStop}%
\end{thebibliography}

%

\end{document}